\documentclass[aps,pra,twocolumn,superscriptaddress,10pt,nofootinbib]{revtex4-2}

\usepackage[utf8]{inputenc}
\usepackage[english]{babel}

\usepackage[T1]{fontenc}
\usepackage{xcolor}
\usepackage{microtype}
\usepackage{enumitem}
\usepackage[normalem]{ulem}
\usepackage[pdftex]{graphicx}
\usepackage{hyperref}
\usepackage{amsmath}
\usepackage{amssymb}
\usepackage{mathtools}
\usepackage{mathrsfs}
\usepackage{multirow}
\usepackage{bbm}
\usepackage{bbold}
\usepackage{braket}
\usepackage[version=4]{mhchem} 
\usepackage[only,llbracket,rrbracket]{stmaryrd}
\usepackage{siunitx}
\usepackage[caption=false]{subfig}
\usepackage{float}
\usepackage{makecell}
\usepackage[percent]{overpic}
\usepackage[capitalize]{cleveref}

\usepackage{nth}

\hypersetup{%
    hypertexnames=false,
    colorlinks=true,
    allcolors=blue,
    unicode=true
}

\newcommand{\phantomsubfloat}[1]{%
    {%
        \captionsetup[subfloat]{farskip=0pt,captionskip=0pt}
        \captionsetup[subfigure]{labelformat=empty}
        \subfloat{#1}
    }
}

\crefname{section}{Sec.}{Sec.}
\crefname{appendix}{App.}{App.}

\newcommand{\ts}{\textsuperscript}

\begin{document}

\title{%
    Boltzmannian state counting for black hole entropy in Causal Set Theory
}
\author{Vid Hom\v sak} 
\thanks{Current affiliation: University of Oxford, Parks Rd, Oxford OX1 3PJ, United Kingdom}
\email{vid.homsak@physics.ox.ac.uk}
\affiliation{Blackett Laboratory, Imperial College London, Prince Consort Rd, London SW7 2AZ, United Kingdom}

\author{Stefano Veroni} 
\thanks{Current affiliation: PlanQC GmbH, Lichtenbergstr. 8, 85748 Garching, Germany}
\email{stefanoveroni00@gmail.com}
\affiliation{Blackett Laboratory, Imperial College London, Prince Consort Rd, London SW7 2AZ, United Kingdom}

\date{\today}

\begin{abstract}
    This paper presents the first numerical study of black hole thermodynamics in Causal Set Theory, focusing on the entropy of a Schwarzschild black hole as embodied in the distribution of proposed horizon molecules.
    To simulate causal sets we created a highly parallelized computational framework in \texttt{C++} which allowed for the generation of causal sets with over a million points, the largest causal sets in a non-conformally flat spacetime to date.
    Our results confirm that the horizon molecules model is consistent with the Bekenstein-Hawking formula up to a dimensionless constant that can be interpreted as the fundamental discreteness scale in the order of a Planck length.
    Furthermore, the molecules are found to straddle the horizon of the black hole to within a few Planck lengths, indicating that entropy lives on the surface of the black hole.
    Finally, possible implications for the information paradox are drawn. 
    In particular, we show how the horizon molecules model could yield a finite black hole temperature cut-off or even prevent full black hole evaporation. 

\end{abstract}

\maketitle

\section{Introduction}

Causal Set Theory (CST) is an approach to the problem of quantum gravity which assumes that continuous spacetime emerges merely as a macroscopic approximation to the fundamental discrete causal structure, a causal set. 
A causal set can be imagined as a network, where the events embedded into spacetime represent the nodes that connect if and only if they are causally related.

If CST aims to be \textit{the} approach to quantum gravity, it is imperative that it gives predictions that agree with our current understanding of the universe and works in the area of physics where quantum gravity would be required.
A perfect example of this is black hole thermodynamics, which arises from the application of Quantum Field Theory on a curved spacetime background~\cite{Hawking1975}.
In this paper, we numerically investigate the horizon molecules model for black hole entropy. 
First developed in Ref.~\cite{Dou2003}, this model assumes that the microstructure of a black hole that gives rise to the Bekenstein-Hawking entropy is formed by molecules of spacetime situated on its horizon.

An algorithm for the simulation of the Schwarzschild black hole in CST was outlined in Ref.~\cite{He2009}.
However, due to the computationally intensive nature of the simulations required to study the behaviour of causal sets in curved spacetime, no extensive numerical studies of CST in the area of black hole thermodynamics exist until now.
To our knowledge, this paper is the first numerical application of causal sets to the area of black hole thermodynamics. 
More generally, before this paper, Ref.~\cite{He2009} was the only numerical study of CST in a non-conformally flat spacetime. 
This paper builds on that original work, creating causal set black holes of cardinality many orders of magnitude larger.

The manuscript is organised as follows.
Section~\ref{sec:background} provides an introduction to the main disciplines involved in our study, namely black hole thermodynamics and causal sets theory.
In \cref{sec:simulating-causets}, we explain how causal sets can be simulated in Schwarzschild spacetime.
Section~\ref{sec: Horizon Molecules} introduces the general theory of horizon molecules, explains their relation to black hole thermodynamics, and outlines the specific characteristics of the ones used in our study.
Section~\ref{Simulation Framework} dives into the highly parallelized \texttt{C++} simulation framework we created to study Schwarzschild causal sets extensively for the first time ever 
and describes how we tested it to make sure the simulations worked as expected.
Lastly, \cref{Results} displays our numerical results, which are then discussed in \cref{Discussion}. 

\section{Background}
\label{sec:background}

\subsection{Black Hole Thermodynamics}\label{BH Thermodynamics}

A black hole is a region of spacetime where gravity is so strong that nothing can escape it. 
The boundary of the black hole is the event horizon: any object crossing the horizon will inevitably keep falling towards the centre, where a singularity is present. This is a region of zero volume where curvature and density diverge, and at which any object's worldline - the object's trajectory in space and time - abruptly interrupts, as if ceasing to exist~\cite{HawkingPenrose1970}. 

Black holes have been shown to be thermal objects endowed with a temperature and entropy~\cite{Bardeen1973, Bekenstein1973}. 
However, as per the no-hair theorem, established via a series of contributions~\cite{Israel1967, Israel1968, Carter1971, Wald1971, Robinson1975, Mazur1982}, black holes appear to be the simplest objects in the universe, fully specified by only 3 observables: mass~\(M\), charge~\(Q\) and angular momentum~\(J\). 
Black hole entropy, which is usually associated with complexity, suggests the existence of hidden degrees of freedom beyond those recognised by the no-hair theorem.

Entropy is a property of statistical mechanics and is usually related to the distribution of
microstates of molecules inside the object of interest. It is an ongoing debate whether the entropy of a black hole counts microstates of the interior of the black hole or whether it is associated to the horizon~\cite{Jacobson_2005}. In this paper we take the attitude that the entropy of a black hole is a property of the horizon~\cite{Sorkin1986}.
 
Heuristically, this is supported by the well-established Bekenstein-Hawking formula~\cite{Bekenstein1972, Bekenstein1973, Hawking1975}
\begin{equation}\label{eq: BH formula}
    S_{BH} = \frac{A}{4 \ell_p^2},
\end{equation}
which states that the entropy of a black hole scales with the \textit{area} $A$ of the event horizon, where $\ell_p = \sqrt{{G\hbar}/{c^3}}$ is the Planck length.

Since the horizon is just a boundary in empty spacetime, this suggests that black hole entropy resides in the microstructure of spacetime itself.
As pioneered by Dou and Sorkin~\cite{Dou2003}, in a discrete spacetime one can connect \lq\lq spacetime atoms'' across the black hole horizon, under certain constraints, to form \textit{horizon molecules}, and then seek in these molecules the origin of the entropy. 

Assuming a large number $N_{mol}$ of independent molecules making up the horizon, each occupying a microstate $n$ with probability $p_n$, one can count them and look at their distribution to measure the Boltzmann entropy of the event horizon
\begin{equation}\label{eq : Boltzmann entropy}
    S = - N_{mol}\sum_n p_n \ln{p_n}.
\end{equation}
Since the Bekenstein-Hawking formula states that entropy scales with the area, we demand that the number of horizon molecules is proportional to the event horizon area, i.e $N_{mol} \propto A$.
This can be studied by simulating causets embedded into Schwarzschild spacetime.

\subsection{Causal Set Theory}\label{CST}

This section gives an overview of CST, explains how causal sets can be approximated by a spacetime continuum and provides the relevant nomenclature.

\begin{figure}[t]
\vspace{4mm}
    \centering
    \phantomsubfloat{\label{fig:causets-rules}}
    \phantomsubfloat{\label{fig: Schwarzschild BH}}
    \begin{overpic}[width=0.95\linewidth]{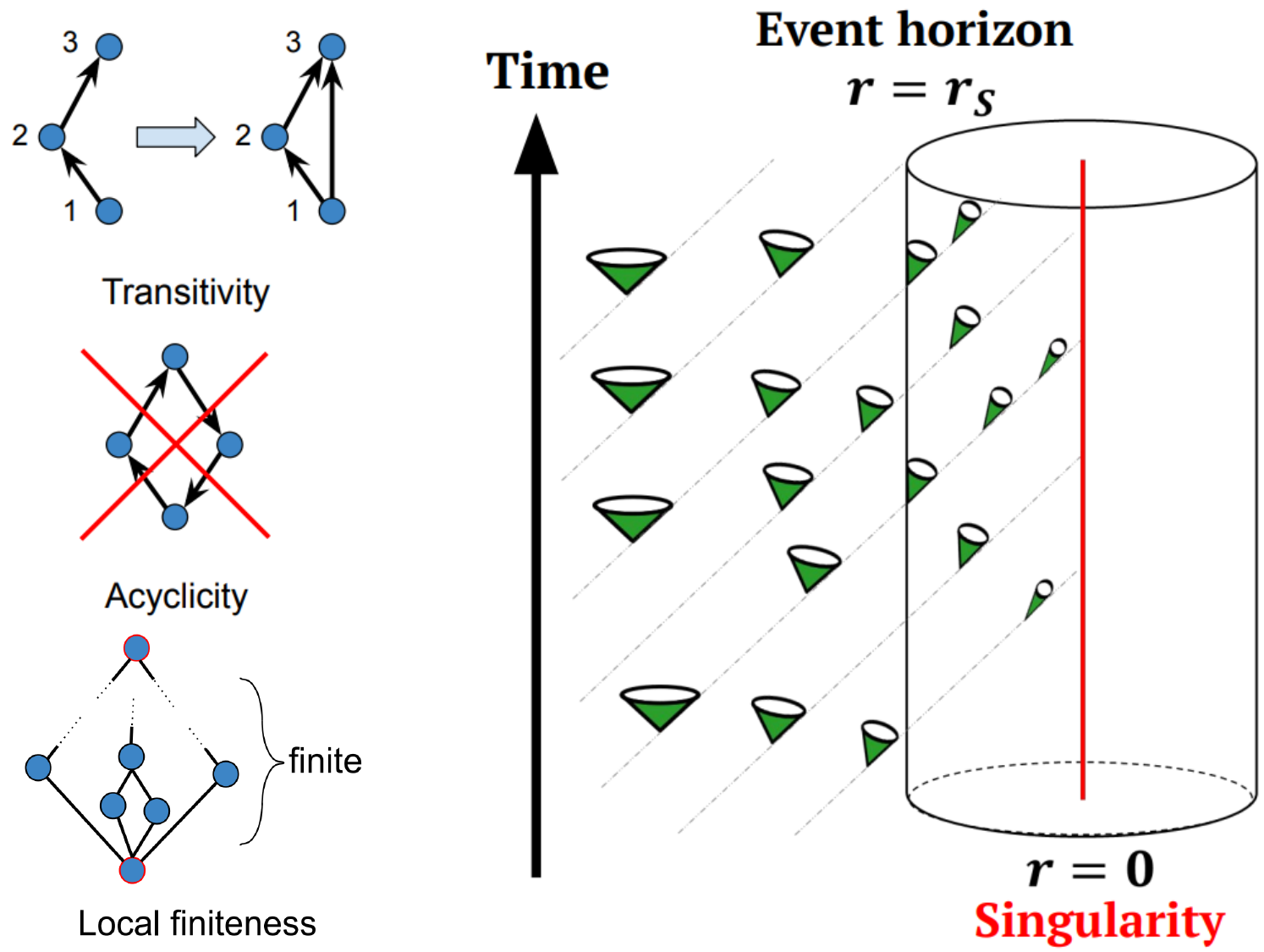}
        \put(-2, 75){(a)}
        \put(33, 75){(b)}
    \end{overpic}
    \caption{
    (a) Causal sets are depicted as network of events. 
    They obey transitivity, acyclicity and local finiteness.
    (b) Simplistic depiction of a Schwarzschild black hole in (2+1)D. 
    Light cones bend as you get closer to the black hole singularity. 
    Once crossing the event horizon at Schwarzschild radius $r_S$, nothing can escape from the black hole, not even light.
    }
\end{figure}

CST began in earnest by Bombelli et al. in 1987~\cite{Bombelli1987}. A causal set (causet) is a discrete partially ordered set of events $C$ with an order relation $\prec$. 
If element $x$ precedes $y$ then $y$ lies in the causal future of $x$, i.e. $x\prec y$. 
Here, $x$ is said to be an \textit{ancestor} ($x$ lies in the \textit{past}) of $y$ and $y$ is a \textit{descendant} (lies in the \textit{future}) of $x$.
Formally, a set of events $C$ with an order relation $\prec$ is defined to be a causal set if it obeys: 
\begin{enumerate}
    \item Transitivity: $x\prec y \land y\prec z \Rightarrow x\prec z$.
    \item Acyclicity: $ x \prec y_1 \prec ...  \prec y_N \neq x$.
    \item Local Finiteness: $|\{z|\ x\prec z \prec y \}|$ is finite.
\end{enumerate}
Transitivity and acyclicity together define the structure to be a partially ordered set. 
Local finiteness asserts the discreteness of spacetime.
One can display a causal set using a Hasse diagram, where the causet takes the form of a network-like structure in which the edges connect the pairs of events that are directly causally related (see \cref{fig:causets-rules}). 
They represent \textit{links}, the irreducible relations between the \textit{parent} $x$ and its \textit{child} $y$. 

Belief in CST stems from the Hawking-King-McCarthy-Malament (HKMM) theorem~\cite{Hawking1976, Malament}, which states that continuous spacetime is fully described by its causal structure up to a local conformal factor, i.e. the causal structure specifies (9/10)th of the metric in 4-dimensional spacetime, only missing an intrinsic measure of volume.
The central conjecture of CST is that this also holds for a discrete spacetime. 
Then, due to local finiteness, a continuum spacetime volume would correspond to a finite number of causet elements which can be counted to provide the missing measure of volume.
Therefore, causal sets are conjectured to be equivalent to spacetime. 

Causet events need to be uniformly distributed across the spacetime volume measure to satisfy the Number $\sim$ Volume correspondence. 
This is possible with a Poissonian relation between the two~\cite{BombelliPhD, Saravani2014}. 
Then, the probability of having $N$ elements in a region of volume $V$ at density $\rho$ is given by
\begin{equation}\label{eq: Poiss probability}
    \text{Poiss}(N)=\frac{(\rho V)^N e^{-\rho V}}{N!}.
\end{equation}
This relation preserves Lorentz invariance, which is imperative for quantum gravity, as no observations so far point towards Lorentz symmetry violation~\cite{Dowker2005, Bombelli2009, Stoyan2013}.

Finally, for an extensive review of CST, we refer the reader to Ref.~\cite{Surya2019}.

\section{Causal Sets in Schwarzschild Spacetime}\label{sec:simulating-causets}

This section describes the computational method used to simulate causal sets in Schwarzschild spacetime.
The process of causet embedding into a chosen (3+1)D spacetime volume region is known as \textit{Poisson sprinkling}, and is outlined as follows:
\begin{enumerate}
    \item Pick the sprinkling density $\rho = \ell^{-4}$, where $\ell$ is the discreteness scale, the average distance between causal set events in 4D spacetime.
    \item Distribute $N\sim\text{Poiss}(\rho V)$ points uniformly, such that the probability of having $N$ points in a volume $V$ is described by \cref{eq: Poiss probability}.
    \item Connect the elements according to the causal structure of the spacetime region.
\end{enumerate}

\subsection{Schwarzschild spacetime} \label{sec:schwarz-spacetime}

In this paper, we study Schwarzschild black holes, the simplest type of black holes. 
They are spherically symmetric, uncharged and have zero angular momentum.
They are only defined by their mass $M$. 
The curvature of the spacetime increases as you move closer to the singularity, which causes the light cones to bend towards the centre of the black hole (see Fig.~\ref{fig: Schwarzschild BH}).
At distance $r_S=2M$ from the centre, the outer edge of the light cone becomes vertical, therefore allowing 
information to flow only in the direction towards the centre of the black hole.
This implies that every event inside the event horizon is causally unrelated to future events outside of it.

Schwarzschild spacetime in 3+1 dimensions is described by a line element which, in Schwarzschild coordinates $(t_S, r, \theta, \phi)$, takes the form of
\begin{equation}
    ds^2 = -\left(1-\frac{2M}{r}\right)dt_S^2 +     \left(1-\frac{2M}{r}\right)^{-1}dr^2 + r^2 d\Omega^2,
\end{equation}
where $d\Omega^2=d\theta^2 + \sin^2{\theta}d\phi^2$. However, Schwarzschild coordinates are problematic as they give a mathematical singularity at $r=r_S$, hence we use Eddington-Finkelstein original 
\footnote{The term ''original'' is used to distinguish them from the better known ingoing (outgoing) coordinates with same name, often denoted as $u$ ($v$). It is motivated by the fact that in the original papers, they use coordinates that resemble those in our paper.}
(EFO) coordinates
\cite{Eddington1924, Finkelstein1958}
\begin{equation}
    \left(t^*=t_S+2M \ln{\left| \frac{r}{2M}-1\right|},\,r,\,\theta,\,\phi\right),    
\end{equation}
which are well-behaved everywhere but at the black hole singularity.
This allows us to rewrite the line element into

\begin{align}
    \label{ds^2 in EF}
     ds^2 =& -\left(1-\frac{2M}{r}\right)dt^{*2} + \frac{4M}{r}dt^{*}dr+   \left(1+\frac{2M}{r}\right)dr^2 \nonumber \\ &+ r^2 d\Omega^2.
\end{align}
The volume element is $dV=r^2 \sin{\theta}\: dt^*\: dr\: d\theta\: d\phi$ and has the symmetry of a (3+1)D cylinder. 

An important feature of the EFO coordinates is that, unlike other sets of coordinates, the constant $t^*$ hypersurface is spacelike everywhere.
This allows us to naturally order events by their time $t^*$ coordinates without further consideration.
To prove this statement, let us define the hypersurface $\Sigma$ as
\begin{equation}
    \Sigma(x^\mu) = t^{*} = \mathrm{const.}
\end{equation}
and its normal vector $n^\mu$ as
\begin{equation}
    n^\mu= \partial^\mu \Sigma.
\end{equation}
Given the inner product 
\begin{equation}
    \label{eq: const-time hypersurface}
    n^\mu n_\mu = g^{\mu\nu}n_\nu n_\mu,
\end{equation}
the hypersurface $\Sigma$ is null if Eq.~(\ref{eq: const-time hypersurface}) yields 0, is timelike if larger than 0 and is spacelike if $n^{\mu}n_{\mu}<0$.
Since the corresponding covector is $n_\mu=(1,0,0,0)$ and the inverse metric in EFO coordinates reads
\begin{equation}
    g^{\mu\nu}=
    \begin{pmatrix}
        -\left(1+\frac{2M}{r}\right) & \frac{2M}{r} & 0 & 0\\
        \frac{2M}{r} & 1-\frac{2M}{r} & 0 & 0 \\
        0 &  0 & \frac{1}{r^2} & 0 \\
        0 & 0 & 0 & \frac{1}{r^2 \sin^2{\theta}}
    \end{pmatrix},
\end{equation}
we thus obtain from Eq.~(\ref{eq: const-time hypersurface}) that $\Sigma$ is indeed spacelike as
\begin{equation}
    \label{eq: t*=const. is spacelike}
    n^\mu n_\mu = -\left(\frac{2M}{r}+1\right) < 0 \; \forall r\in\mathbb{R^+}.
\end{equation}

\subsection{Schwarzschild causal structure}

Until now, \textit{all} extensive CST numerical studies have been done in flat Minkowski spacetime~\cite{Surya2019}, where it is rather trivial to obtain the causal relations amongst the events: 
to determine whether event $E_1 = (t_1,\Vec{x}_1)$ is causally related to event $E_2=(t_2,\Vec{x}_2)$, one
simply demands a null or positive spacetime interval $\Delta s^2 = (\Delta t)^2 - |\Delta\Vec{x}|^2$.

On the other hand, things become drastically more complicated in spacetimes that are not conformally flat, such as the one of the Schwarzschild black hole. 
In that case, integration of infinitesimal invariant distance $ds$ along all possible paths from $E_1$ to $E_2$ is in principle required to determine the causality, which is computationally very expensive.

\begin{figure*}[]
    \vspace{1mm}
     \begin{overpic}[width=0.9\textwidth]{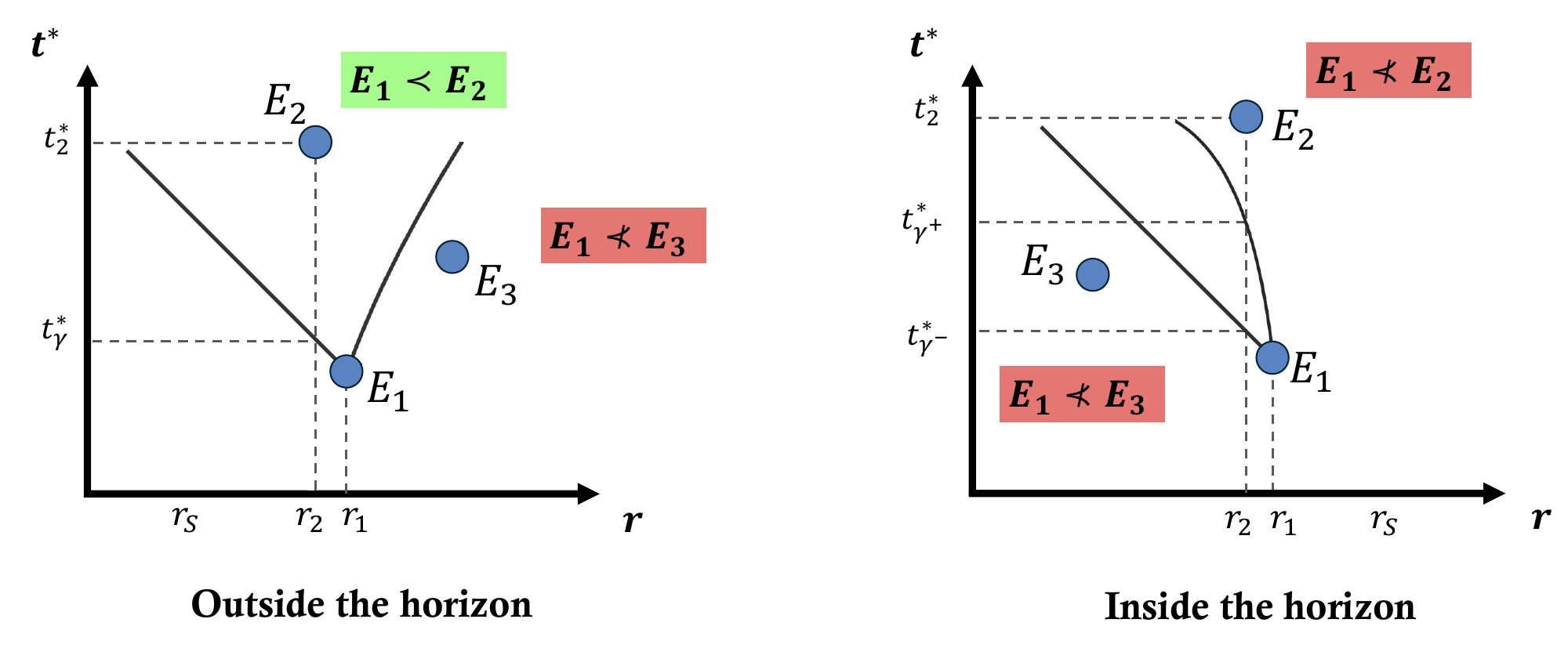}
        \put(-2, 40){(a)}
        \put(54, 40){(b)}
    \end{overpic}
     \vspace{-1mm}
    \caption{
        Causal relations in Schwarzschild spacetime in (1+1)D, outside (a) and inside (b) the event horizon. 
        Specifically, null geodesics departing from event $E_1$ are shown and used to infer its causal relations with other events.
        We define $t^*_{\gamma^-}$ as the time at which the earliest null geodesic departing from $E_1$ reaches the spatial position of $E_2$.
        Similarly, $t^*_{\gamma^+}$ is the latest.
        a) Outside the event horizon, only the earliest $t^*_\gamma$ exists, and it is sufficient to check whether this occurs before $t^*_2$, i.e if $t^*_{\gamma}\leq t^*_2$. 
        b) Inside the event horizon, one needs to check whether $t^*_2\in [t^*_{\gamma^-}, t^*_{\gamma^+}]$, correcting the method given in Ref.~\cite{He2009}.
        }
        \label{fig: Schwarzschild_causality}
\end{figure*}

However, a better algorithm was introduced by He and Rideout~\cite{He2009}. 
For a general pair of events $E_1 = (t_1^*,r_1,\theta_1,\phi_1)$ and $E_2 = (t_2^*,r_2,\theta_2,\phi_2)$, the method determines their causal relation by checking whether $E_2$ lies inside the future light cone of $E_1$.
Since the boundaries of light cones are determined by null curves, the method distils itself into finding the null geodesics emanating from $E_1$, and comparing $t^*_2$ 
with the times $t^*_\gamma$ when null geodesics reach the spatial position of $E_2$ (see Fig. \ref{fig: Schwarzschild_causality}).

However, Ref.~\cite{He2009} assumes that a line of constant spatial position is a possible trajectory everywhere, which is wrong. Due to the bending of the light cones, no object can remain at the same spatial position inside the horizon. 
The corrected procedure is derived in the following subsections, first for the simple (1+1) dimensional spacetime, then in dimensions up to (3+1).

\subsubsection{Causality in (1+1)D Schwarzschild spacetime}

Consider two events $E_1= (t^*_1,r_1)$ and $E_2=(t^*_2,r_2)$. The line element from Eq.~(\ref{ds^2 in EF}) in 1+1 dimensions yields
\begin{equation}
    ds^2 = -{dt^*}^2 + dr^2 + \frac{2M}{r} (dt^*+dr)^2,
\end{equation}
which can be rewritten into
\begin{equation}
    ds^2 = (dt^*+dr)\left[\left(\frac{2M}{r}-1\right)dt^* 
                        + \left(\frac{2M}{r}+1\right)dr\right].
\end{equation}
Setting $ds^2=0$ yields the two radial null geodesics (see~\cref{fig: Schwarzschild_causality}).
The first one gives
\begin{equation}
    t^*_2 - t^*_1 = r_1 - r_2,
\end{equation}
describing an always ingoing null geodesic (left/lower straight line in~\cref{fig: Schwarzschild_causality}) while the second one gives
\begin{equation}
    \label{eq: EqNull2D_up}
          t^*_2 - t^*_1 = \left[r + 4M \ln{(r-2M)}\right]_{r_1}^{r_2}.
\end{equation}
describing the other geodesic, which is ingoing inside the horizon and outgoing outside of it (upper curved in~\cref{fig: Schwarzschild_causality}).

Therefore, we are left with the following possible scenarios.
\begin{enumerate}
    \item If $E_1$ is outside of the event horizon, then the necessary and sufficient condition is 
    \begin{equation}
    t^*_2 - t^*_1 \geq 
    \begin{cases}
        r_1 - r_2, & \textrm{if} \;\; r_2 \leq r_1  \\
        r_2 - r_1 + 4M \ln{\left(\frac{r_2-2M}{r_1-2M}\right)} \; & \textrm{if} \;\; r_2 > r_1
        \label{eq: 2D Schwarzschild lower bound}
    \end{cases}.
\end{equation}
    \item If $E_1$ is inside the event horizon and $r_2>r_1$, they are trivially causally unrelated.
    \item If both events are inside the event horizon and $r_1 \geq r_2$, then the two events are connected if and only if
    \begin{equation}\label{eq: 2D Schwarzschild upper bound}
    r_1 - r_2 \leq
    t_2^* - t^*_1 \leq
    r_2 - r_1 + 4M \ln{\left(\frac{2M-r_2}{2M-r_1}\right)}.
    \end{equation} 
\end{enumerate}
The third point corrects Ref.~\cite{He2009}, where it is stated that \lq\lq\textit{for both events inside the horizon, the necessary and sufficient condition for their causal relation is still $t^*_2 \geq t^*_1 + r_1 - r_2$}'', which is wrong as it neglects the upper bound from Eq.~(\ref{eq: EqNull2D_up}).

\subsubsection{Causality in (3+1)D Schwarzschild spacetime}\label{sec: causality in 3+1 D spacetime}

Noting that the method always involves only two events $E_1 = (t_1^*,r_1,\theta_1,\phi_1)$ and $E_2 = (t_2^*,r_2,\theta_2,\phi_2)$ simultaneously, we can simplify our calculations in (3+1)D EFO coordinates by rotating the spatial part of the coordinate system to an equatorial plane containing both of the events, such that $\theta'_1=\theta'_2=\pi/2$ and $\phi'_1=0$.
{As the equatorial plane is totally geodesic, we can effectively neglect $\theta$ and rewrite their coordinates as} 
\begin{align}
    & E_1 = \left(t_1^*, r_1, 0\right),\\
    & E_2 = \left(t_2^*, r_2, \arccos{\left(\sin{\theta_1}\sin{\theta_2}\cos{\Delta\phi} + \cos{\theta_1}\cos{\theta_2}\right)} \right), \label{eq: E_2 after rotation}
\end{align}
reducing the problem to 2+1 dimensions.
Compared to the radial (1+1)D case, in (3+1)D the procedure to determine the causal relations becomes more involved since it is impossible to solve for null geodesics analytically. 

Given an affine parameter $\lambda$, they satisfy 
\begin{align} \label{4D null geodesic}
     &\left(1-\frac{2M}{r}\right)
           \left(\frac{dr}{d\lambda}\right)^2
        -\left(1-\frac{2M}{r}\right)
           \left(\frac{dt^*}{d\lambda}\right)^2
        + r^2 \left(\frac{d\phi}{d\lambda}\right)^2  \nonumber \\ &= 0,
\end{align}
and since the line element in~\cref{ds^2 in EF} is independent of $t^*$ and $\phi$, the following are conserved quantities
\begin{align}
    E &= \left(1-\frac{2M}{r}\right) \frac{dt^*}{d\lambda},\\
    L &= r^2 \frac{d\phi}{d\lambda}.
\end{align}
Moreover, we can rewrite the derivatives as
\begin{align} \label{drdphi}
    \frac{dr}{d\lambda} &= \frac{dr}{d\phi} \frac{d\phi}{d\lambda}
    = \frac{L}{r^2} \frac{dr}{d\phi},\\
    \frac{dt^*}{d\lambda} &= \frac{dt^*}{d\phi} \frac{d\phi}{d\lambda}
    = \frac{L}{r^2} \frac{dt^*}{d\phi},
\end{align}
to get rid of $\lambda$ and re-express $dr/d\phi$ in terms of $u=1/r$, which is computationally advantageous as it leads to expressions with lower powers, such that
\begin{equation}
    \frac{dr}{d\phi} = u^{-2} \frac{du}{d\phi}.
\end{equation}
Furthermore, we can define the ratio
\begin{equation}
    \eta = \frac{E}{L},
\end{equation}
which is necessarily finite since $L=0$ describes the (1+1)D case with $\Delta\phi=0$.
Finally, inserting previous equations into Eq.~(\ref{4D null geodesic}), we obtain the two differential equations
\begin{align} 
    \label{dphidu}
        \frac{d\phi}{du} &= \pm_{du} \left(\eta^2 + 2Mu^3 - u^2 \right)^{-1/2}\\
         \frac{dt^*}{du} &= \frac{1}{u^2(2Mu-1)} 
    \left[
    \frac{\pm_{du}\eta}{\sqrt{\eta^2 + u^2(2Mu-1)} } + 2Mu \right]
    \label{dt*du}
\end{align}
where
\begin{equation}
    \pm_{du} = \mathrm{sign}(u_2 - u_1)
\end{equation}
depends on the order in which the two events were initially chosen.

Since the initial and final spatial boundary conditions $(u,\phi)$ are known, we can fit Eq.~(\ref{dphidu}) for $\eta^2$ to obtain $\pm\eta$ values and then use them to integrate Eq.~(\ref{dt*du}). 
This gives us two values $t^*_+$ and $t^*_-$, corresponding to the times when each null geodesic reaches the spatial coordinate of $E_2$.

In Appendix \ref{app:4Dcausality}, we prove explicitly the conditions for causality. 
We show that in the case that $E_1$ is outside of the event horizon, only one of the times $\{t^*_-,t^*_+\}$ is physically meaningful, $t^*_-$.
Then, $t_2>t^*_-$ is the sufficient and necessary result to determine that $E_2$ lies in the causal future of $E_1$, i.e that $E_1\prec E_2$.
On the other hand, if both points lie inside the horizon such that $r_S>r_1>r_2$, $E_1\prec E_2$ if and only if $t_2 \in [t^*_-,t^*_+]$. 
In fact, the bending of light cones (see Fig. \ref{fig: Schwarzschild_causality}), poses an upper time bound.
Finally, we also show that the value of $\phi$ should be restricted to lie in $[0, 2\pi]$. 

\subsubsection{Sufficient conditions for determining the causality}

It is worth noting that a large number of event pairs in (3+1)D does not require the general approach with integration and fitting for $\eta^2$ to determine the causal structure of the Schwarzschild spacetime.

Instead, there exists a set of simple sufficient but not necessary conditions that we check, for each pair of events, before applying the generic method outlined above to determine whether the points are causally related or not. 
This speeds up the procedure significantly since the checks are computationally very inexpensive compared to the fitting and integration.
To get the full list and the derivations of the conditions, refer to Sec. 2.3 in Ref.~\cite{He2009}.

\section{Horizon molecules}\label{sec: Horizon Molecules}

This section defines horizon molecules, determines the coefficient for the scaling of number of molecules with the area of the black hole horizon and calculates the curvature corrections in Schwarzschild spacetime.

In this paper, we investigated a new type of horizon molecule, the $\Lambda_n$ molecule, which will be discussed in detail in the following subsection.

\begin{figure*}
    \includegraphics[width=\linewidth]{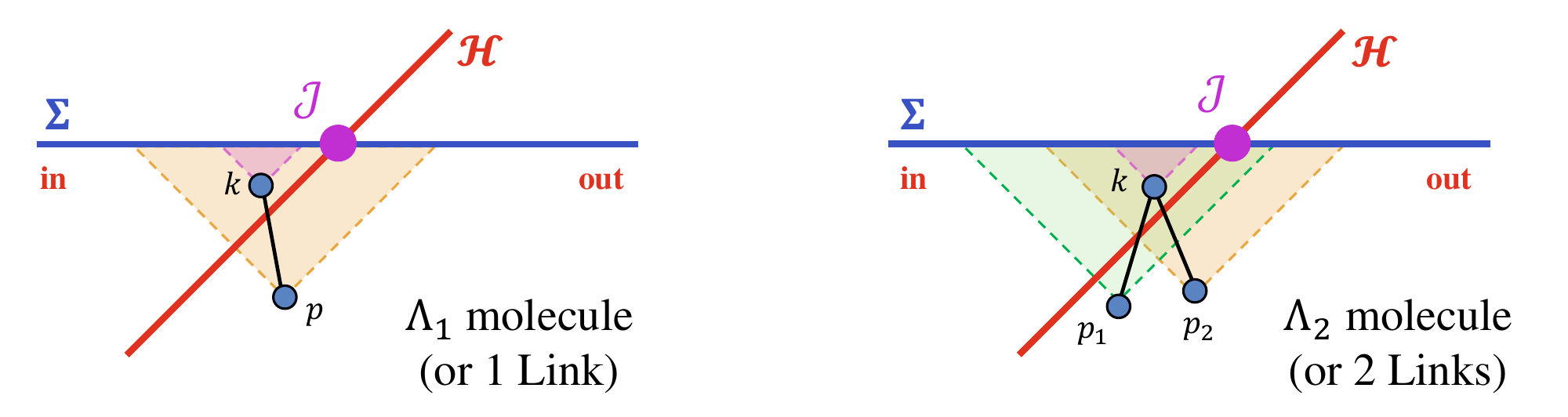}
    \vspace{-5mm}
    \caption{Schematic representation of $\Lambda_n$ and Link horizon molecules. They consist of connected points inside and outside the event horizon $\mathcal{H}$, satisfying certain constraints. 
    The spacelike hypersurface $\Sigma: \: t^*=\mathrm{constant}$ defines the time of the entropy measurement. 
    $\Lambda_n$ molecules are defined by having an element $k$ inside the horizon which is maximal with respect to $\Sigma$ and connects to $n$ elements outside the horizon that only contain $k$ in their causal future below $\Sigma$.
    A single $\Lambda_n$ molecule is equivalent to $n$ Links. 
    }
    \label{fig:molecules_schematic}
\end{figure*}

In general, a horizon molecule consists of one point inside the event horizon $\mathcal{H}$ connected to one or more points outside it via links and lies below some $t^*=constant$ hypersurface $\Sigma$. 
In EFO coordinates, $\Sigma$ is spacelike everywhere as shown in Eq.~(\ref{eq: t*=const. is spacelike}), which allows us to think of $t^*$ coordinate as time.
Since entropy is only defined as a function of time, $\Sigma$ signifies the moment when entropy is \lq\lq measured''. 
Therefore, one can count the number of molecules, study their distribution, and obtain results for the proportionality factor for $N_{mol}\propto A$ in Eq.~(\ref{eq : Boltzmann entropy}).

On the other hand, scaling can be derived analytically for certain molecules. 
Knowing that the probability that a region $\Delta R$ of spacetime with volume $V$ contains $N$ events is given by the Poisson distribution as in Eq.~(\ref{eq: Poiss probability}),
we can calculate the expected number of horizon molecules and compare it with the numerical result. 

Analytic studies~\cite{Barton2019, Dou2003} assume that at sufficiently high sprinkling density, molecules are localised in space and time.
This allows us to work in flat approximation with Rindler coordinates,
where the event horizon and the constant time hypersurface are given by
\begin{equation} \label{eq: Rindler horizon}
            \mathcal{H}: t=x, \quad \Sigma: t=\mathrm{const.}
\end{equation}
This flat-space representation is used to display $\Lambda_n$ and Link horizon molecules, described in the following subsection, in Fig.~\ref{fig:molecules_schematic}.

\subsection{\texorpdfstring{$\Lambda_n$}{Lambda} and Link molecules}
\label{sect: nlambda mulecules}

The $\Lambda_n$ molecule is given by a maximal element $k$ inside the event horizon that lies in the causal future of $n$ mutually disconnected elements $\{p_n\}$ outside of it (see Fig. \ref{fig:molecules_schematic}),
which can only contain $k$ in their causal future below $\Sigma$.
Unfortunately, we cannot derive the analytic expression for the average expected number of $\Lambda_n$ molecules.

However, we can derive the analytical expression for Link molecules, which are closely related to $\Lambda_n$. They are links crossing the horizon whose past element is maximal but one with respect to $\Sigma$ 
\footnote{Upper-case Link denotes the horizon molecule, whereas lower-case link the nearest-neighbour relation between causet events.}.
The analytic result for the average expected number of Links $\langle N_{L} \rangle$ can then be compared with the numerical result obtained by weighted counting of $\Lambda_n$ molecules, where each $\Lambda_n$ contributes $n$ Links.

The expression for $\langle N_L \rangle$ has been computed by Barton et al.~\cite{Barton2019}. Here we sketch the calculation.
Given the discreteness density $\rho$, the probability that a point $p$ outside the horizon in the region $\delta R_p$ connects only to one maximal future point inside the horizon is given by
\begin{align}
    P_{L}(\delta R_p)
    &= P(1\;\mathrm{in}\;\delta R_p) 
    \cdot P(1\;\mathrm{in}\;I^+_{in}(p)) \\
    &= \rho^2 \delta V_p V^+_{in}(p) e^{-\rho V^+_{in}(p)}. \label{eq: prob for link}
\end{align}
$V^+(p)$ is the volume of the future light cone of $p$ capped by the $\Sigma$ hypersurface (coloured orange in the $\Lambda_1$ schematic in \cref{fig:molecules_schematic}) made of $V^+_{in}(p)$ and $V^+_{out}(p)$ partial volumes, inside and outside the horizon respectively, and represents $I^+(p)$, the causal future of $p$. 
We denote $\mathcal{J}$ as the intersection of the event horizon $\mathcal{H}$ and hypersurface $\Sigma$, which represents the event horizon at time $t_\Sigma$, and is just a point in 1+1 dimensions.

Integrating over all spacetime regions $\delta R_p$ in the past of the horizon and hypersurface intersection $I^-(\mathcal{J})$, yields the expected number of Links
\begin{equation}
    \label{eq: expected nb of links general expression}
    \langle N_{L} \rangle = 
    \rho^2 \int_{I^-(\mathcal{J})} 
    dV_p\;V^+_{in}(p)\;e^{-\rho V^+_{in}(p)}.
\end{equation}
Noting that we are working with Rindler horizon in (3+1)D flat spacetime, analytic expressions for the volumes can be found.
The average expected number of Links for the horizon area $A$ in the flat approximation is then given by
\begin{equation}
    \langle N_{L}^{flat} \rangle = \frac{\sqrt{3}}{10} \sqrt{\rho} A.
\end{equation}
 In terms of the discreteness scale $\ell = \rho^{-1/4}$, this reads
\begin{equation}\label{eq: avg expected nb of links 0order}
    \langle N_{L}^{flat}\rangle =  a_{L}^{(0)} \frac{A}{\ell^2} = \frac{\sqrt{3}}{10}\frac{A}{\ell^2}.
\end{equation}

\subsection{Curvature corrections in Schwarzschild spacetime}\label{sect: curv corrections}

\begin{figure*}
    \includegraphics[width=0.95\textwidth]{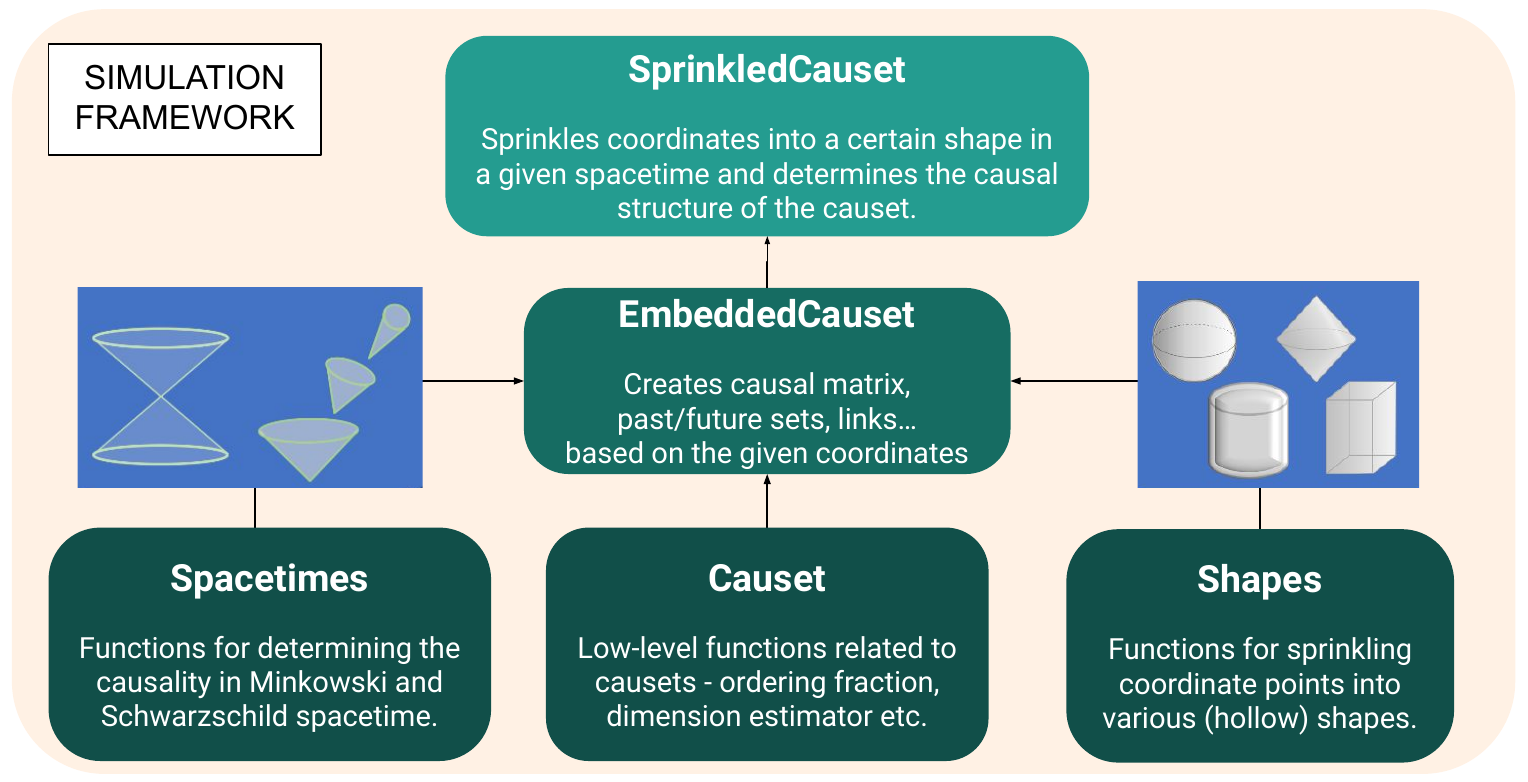}
    \vspace{-4mm}
    \caption{The \texttt{C++} simulation framework available on \href{https://github.com/vidh2000/MSci_Schwarzschild_Causets}{\texttt{GitHub}}. It allows for an efficient generation of causal sets in Minkowski and Schwarzschild spacetime.}
    \label{fig: simulation framework}
\end{figure*}

In general, the average number of molecules, for any horizon $\mathcal{H}$, is given by an infinite series
\begin{equation}
    \label{eq:nmolecules-expansion}
    \langle N \rangle = a \frac{A}{\ell^2} =  
         \left[a^{(0)} + a^{(1)}(\mathcal{H}, \Sigma) \frac{\ell}{A} + \mathcal{O}\left(\frac{\ell^2}{A}\right)\right] \frac{A}{\ell^2}.
\end{equation}
To go beyond the flat spacetime approximation $a^{(0)}$ and estimate the 1st order curvature correction $a^{(1)}$, which depends on the structure of \(\mathcal{H}\) and \(\Sigma\), we follow the procedure described by Sec. 5 in Barton et. al~\cite{Barton2019}.

The general expression for the 1st order correction of Links in terms of the trace $K$ of the extrinsic curvature of $\Sigma$, its component $K_m$ tangential to $\Sigma$ and orthogonal to $\mathcal{J}$, and the null expansion $\vartheta$ of the horizon, all evaluated on $\mathcal{J}$ in 3+1 dimensions, is given by
\begin{equation}
    \label{eq: curv scale correction expression}
    \frac{a_L^{(1)}}{\sqrt{A}} \approx -  (0.036 K + 0.088 K_m + 0.021\vartheta).
\end{equation}
In Appendix \ref{appendix: curv corrections}, we compute $a_L^{(1)}$ for the specific case of the Schwarzschild spacetime. This gives
\begin{equation}
    \label{eq: link curv correction}
     a_L^{(1)} \approx  -0.0558.
\end{equation}
Therefore, the average expected number of Links in Schwarzschild spacetime is 
\begin{equation}
     \label{eq: Nlink corrected for curv}
    \langle N_{L}^{BH}\rangle = a_{L}\frac{A}{\ell^2} = 
    a_L^{(0)}\frac{A}{\ell^2} \left( 1 -0.322 \frac{\ell}{\sqrt{A}} + \mathcal{O}(\ell^2) \right).
\end{equation}
The correction becomes negligible when $A$ increases to the size of ordinary black holes since $\ell\sim \ell_p$ is infinitesimal.

\section{Simulation Framework}\label{Simulation Framework}

This section describes the simulation framework we constructed to efficiently simulate causal sets embedded into Schwarzschild spacetime and the tests we implemented to ensure it works as intended.

\begin{figure*}
    \phantomsubfloat{\label{fig:test-poisson}}
    \phantomsubfloat{\label{fig:test-bh-causal}}
    \includegraphics[width=\textwidth]{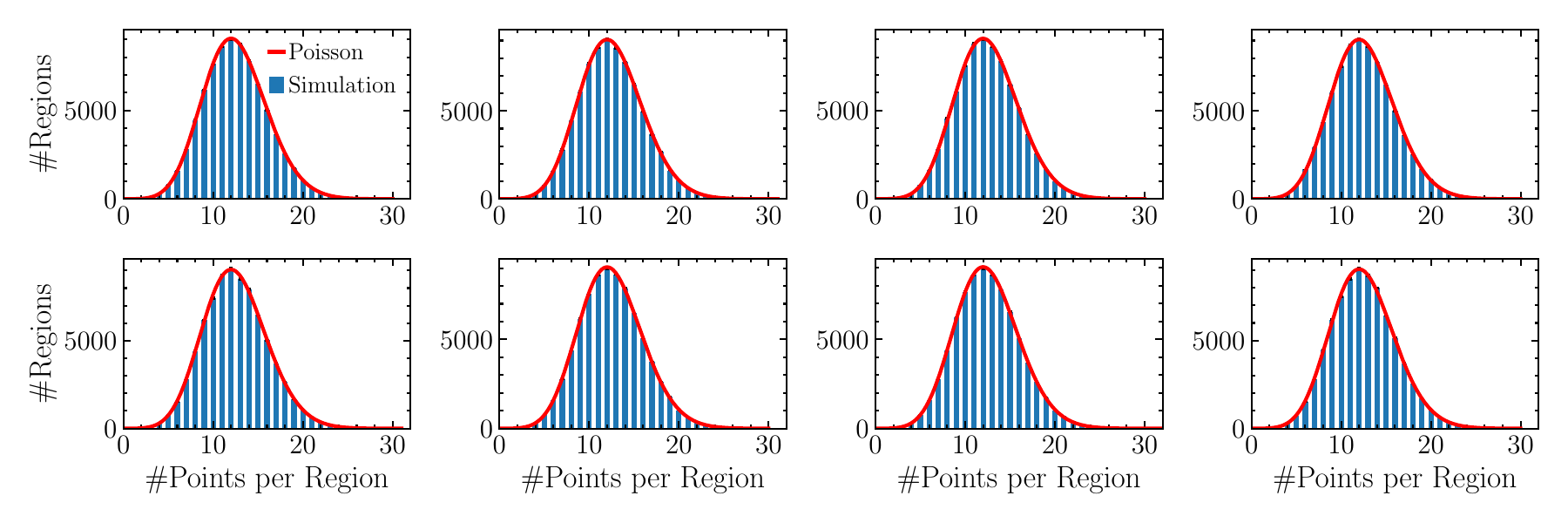}  
    \includegraphics[width=0.32\textwidth]{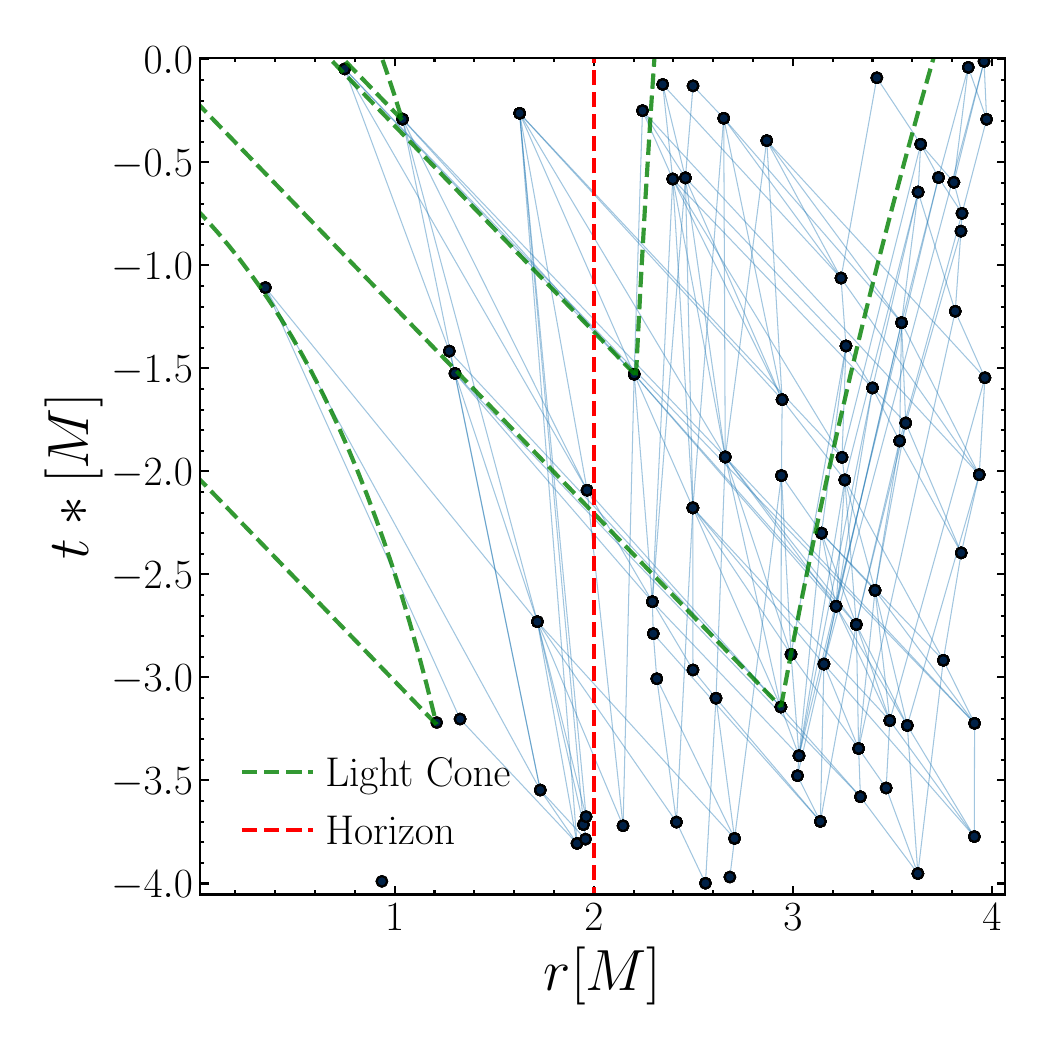}
     \includegraphics[width=0.32\textwidth]{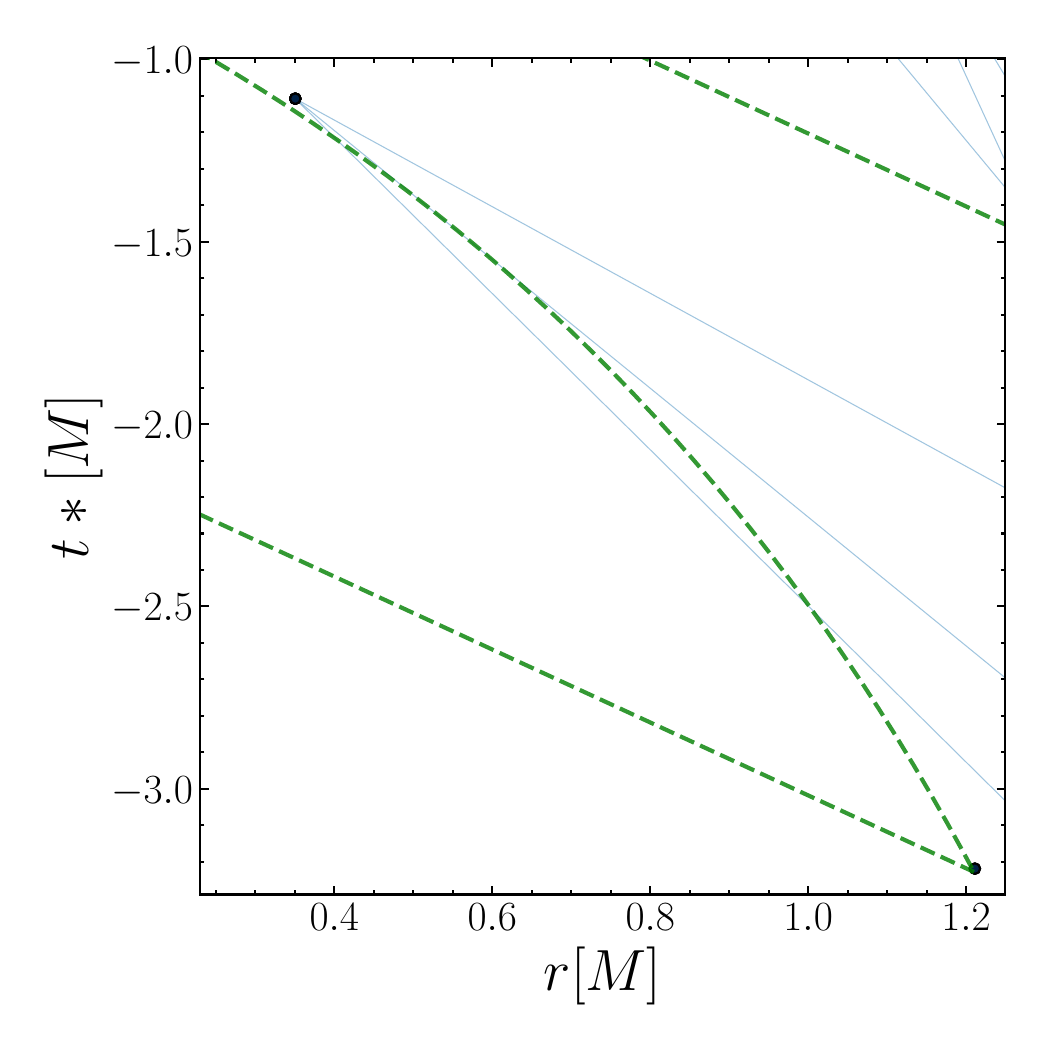}
    \begin{overpic}[width=0.32\textwidth]{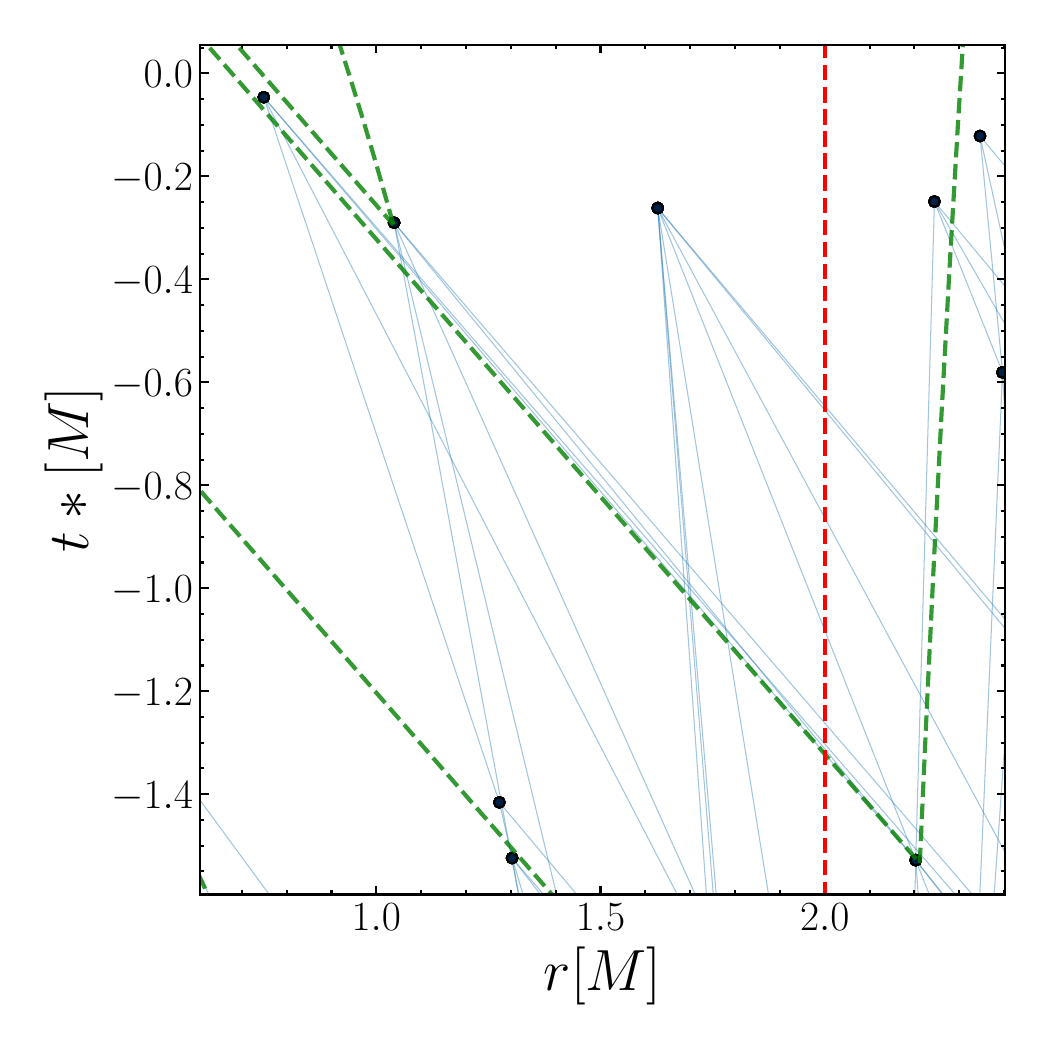}
        \put(-206, 200){(a)}
        \put(-206, 97){(b)}
        \put(-102, 97){(c)}
        \put(-1, 97){(d)}
    \end{overpic}
    \caption{{Main tests for causet generation in Schwarzschild spacetime. 
    (a) Test of the Poisson distribution of events by sampling the number of points in regions of equal volume over 100 causet realizations with the cardinality of $10\,000$ each. 
    The first row refers to sprinkling into a filled cylinder, whereas the second to a hollow cylinder. 
    The columns refer to cutting regions of spacetime limited in $t^*$, $r$, $\theta$ and $\phi$, respectively. 
    Points are indeed distributed following a Poissonian distribution. 
    (b) 2D projection of a very thin slice ($0.25^\circ$) of a 3D black hole causet. 
    (c) and (d) highlight some details of (b), showing that even when events are extremely close to the null geodesics, they are only connected if and only if inside the light cone, as it should be.}}
    \label{fig:main-tests-schwarz}
\end{figure*}

It is important to note that other than the work done in Ref.~\cite{He2009}, which laid out the foundations, there exist no extensive studies of Schwarzschild causal sets. 
Particularly this is because the procedure used to determine the causal structure of Schwarzschild spacetime, as described in \cref{sec:simulating-causets}, is extremely computationally expensive since for a generic pair of events minimization and integration need to be performed. 
Based on our experience, generating Schwarzschild causets is approximately 200 times slower compared to flat spacetime, which explains why other causet numerical studies did not attempt it.

While we initially used the causal set package written by Minz~\cite{Minz}, we realised that Python is too slow and decided to create our simulation framework in \texttt{C++}, due to its speed. 
The package is publicly available on \href{https://github.com/vidh2000/MSci_Schwarzschild_Causets}{\texttt{GitHub}} and allows the user to simulate causal sets in regions of different shapes in both Minkowski and Schwarzschild spacetime for the dimensions $D=2,3,4$. 
It furthermore contains functions to analyse horizon molecules, estimate dimensions of causal sets, plot them and more. 
It is the most efficient causet package publicly available to our knowledge.

In general, during the causet generation, our program only saves events' coordinates and the causal matrix $C$,  where $C_{ij}=1$ if $i\prec j$ and $C_{ij}=0$.
The framework also allows for storing the sets of points in the past and/or future sets and corresponding sets of past/future links for each element as they could be preferable for certain applications.
This drastically speeds up the process while simultaneously reducing memory usage. 
The latter in the end turned out to be an important bottleneck of the simulation, since the realization of a causet with $N = 10^6$ required up to 4 TB of RAM.

Our code uses CPU parallelization as much as possible to speed up the simulations.  Importantly, we used dynamical (unordered) parallelization by \href{https://www.openmp.org/}{\texttt{OpenMP}}, which allows for threads that have already completed their tasks to take on new ones, not requiring to wait for the slowest thread to finish.
In our simulations, we achieved maximum causet cardinality $N = 1,011,423$, by far the largest recorded causal set in curved space, which allowed us to obtain sufficient statistics. 
Overall, the estimated speed-up of the causal set generation in the parallelized \texttt{C++} framework on 256 cores using Imperial College's HPC cluster compared to the non-parallelized Python setup on a personal laptop, using the same generation algorithms, comes out to be approximately a factor of 15000.

\subsection{Testing the simulation framework}\label{Testing}

It is crucial to ensure that the simulation framework works as intended.
Therefore, we implemented various checks to test the validity of our simulations. 
This section describes the checks relevant to the purposes of this article, whereas other testing
of the framework is described in Appendix \ref{appendix: Extra simulation framework tests}.

As per \cref{sec:simulating-causets}, the embedding of a causet in spacetime relies on two ingredients: a Poissonian correspondence between the number of events and spacetime volume, and a correct derivation of causal relations. 
To assess whether events were correctly distributed, we divided the simulated embedding into regions of equal spacetime volumes and counted the number of events in each region.
The distribution of regions with a specific number of events indeed reproduced a Poisson distribution, both for a full and hollow cylinder of spacetime (see \cref{fig:test-poisson}).

Due to the novelty of our simulations, we lacked numerical results to benchmark our code with. In particular, to test the correctness of the causal relation. 
However, we could visually assess the correctness of 
(1+1)D simulations by checking that causal relation only occur within light cones, for which analytic solutions are given by Eq.~(\ref{eq: 2D Schwarzschild lower bound}) and Eq.~(\ref{eq: 2D Schwarzschild upper bound}).
In the same way, we could test (2+1)D simulations in the 2-dimensional limit. 
This was done by sprinkling causets in a very thin slice of a cylinder and projecting the results on a plane (see \cref{fig:test-bh-causal}).
Moreover, the number density scales with $r$, as expected for (2+1)D spacetime.
Note that this also assesses the correctness of the 4D simulations, as the methods used to simulate black hole causets in 4 and 3 dimensions are exactly the same, due to the symmetry about $\theta = \pi/2$. 

\section{Results}\label{Results}

This section describes the results we obtained on the entropy and the properties of the horizon molecules.

We simulated causets in Schwarzschild spacetime with density $\rho=\ell^{-4}=5000$ and cardinality up to $\sim 10^6$ events to count the number of $\Lambda_n$ molecules and study their distribution. Each data point in the following plots is averaged over 200 realizations.

\begin{figure*}    
    \includegraphics[width=0.49\textwidth]{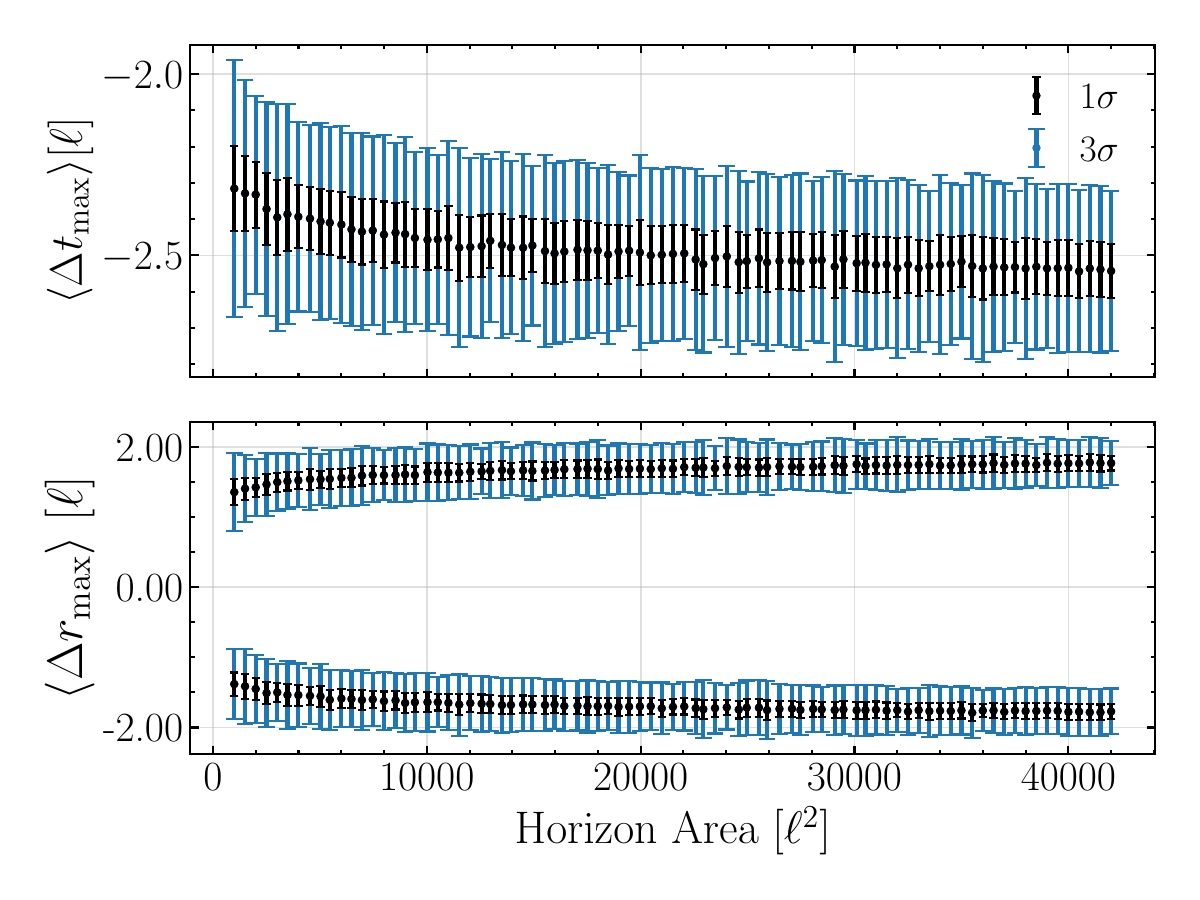}
    \begin{overpic}[width=0.49\textwidth]{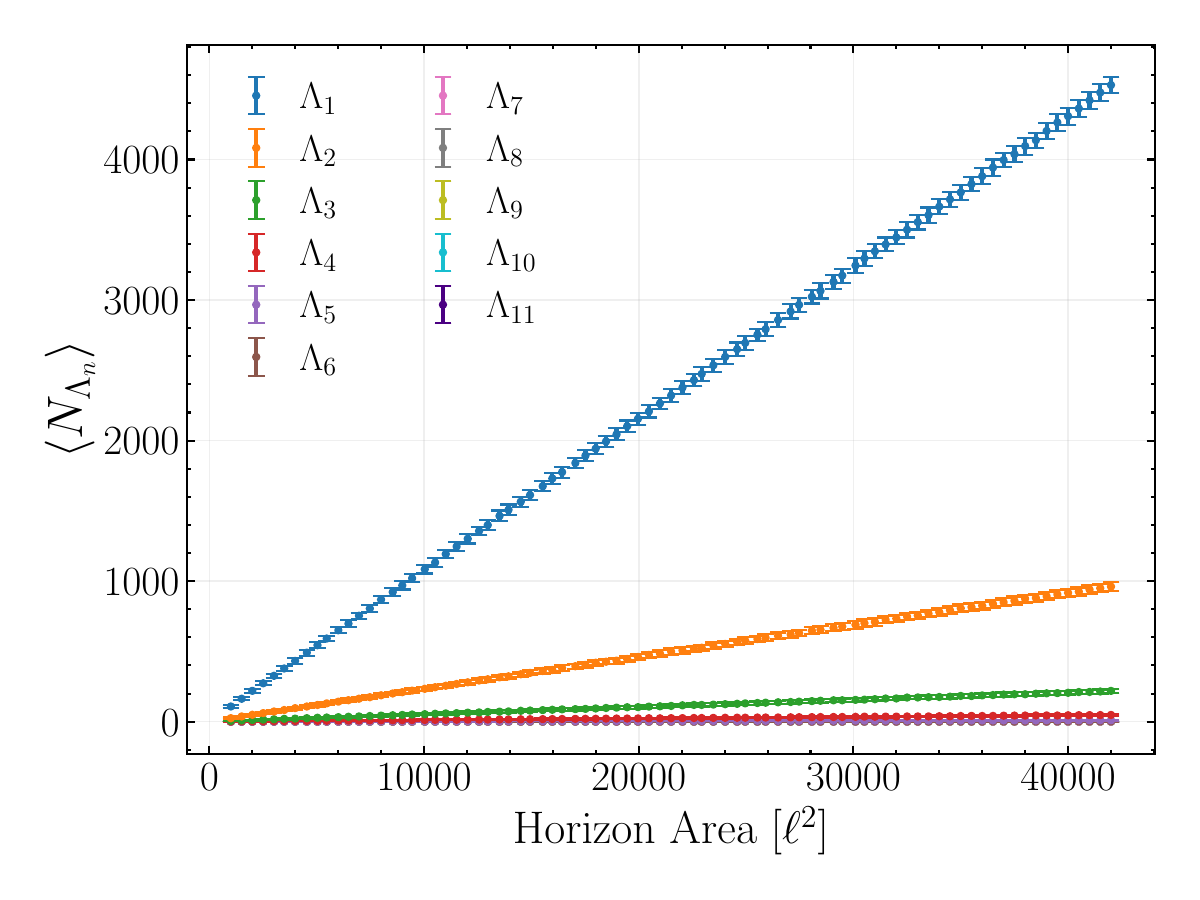}  
        \put(-100, 73){(a)}
        \put(0, 73){(b)}
    \end{overpic}
    \includegraphics[width=0.57\textwidth]{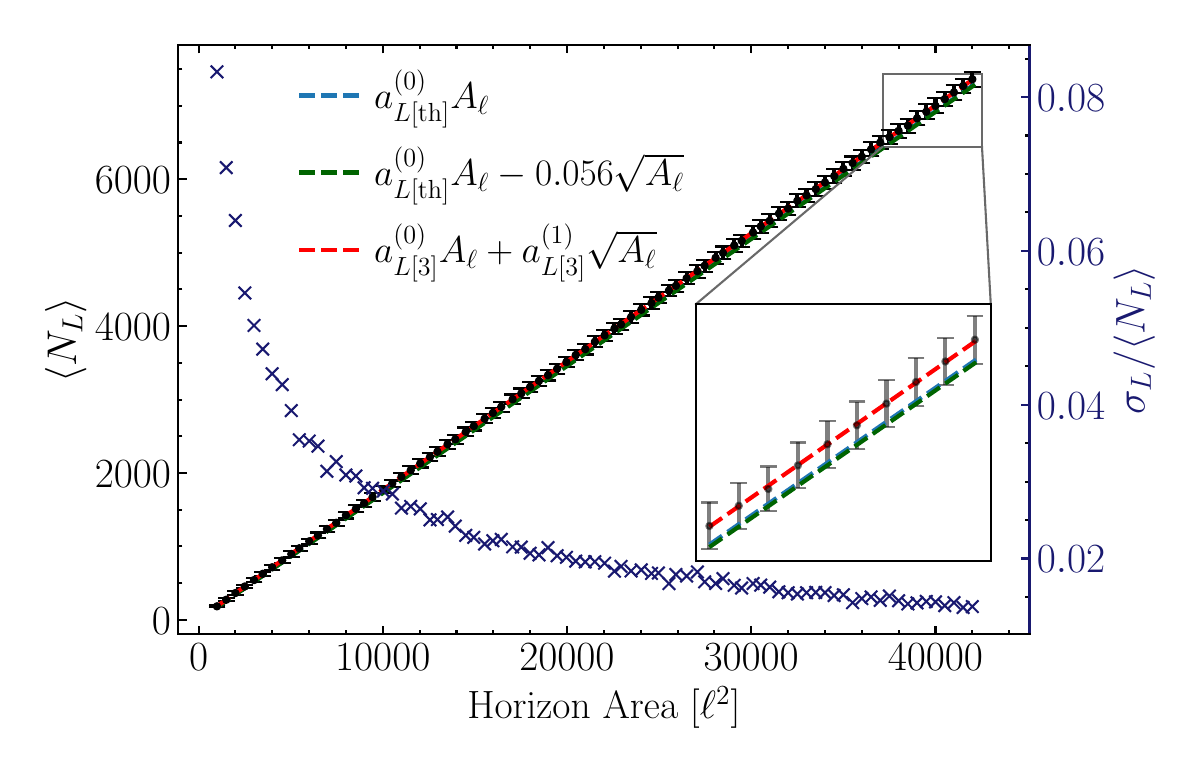}
    \begin{overpic}[width=0.42\textwidth]{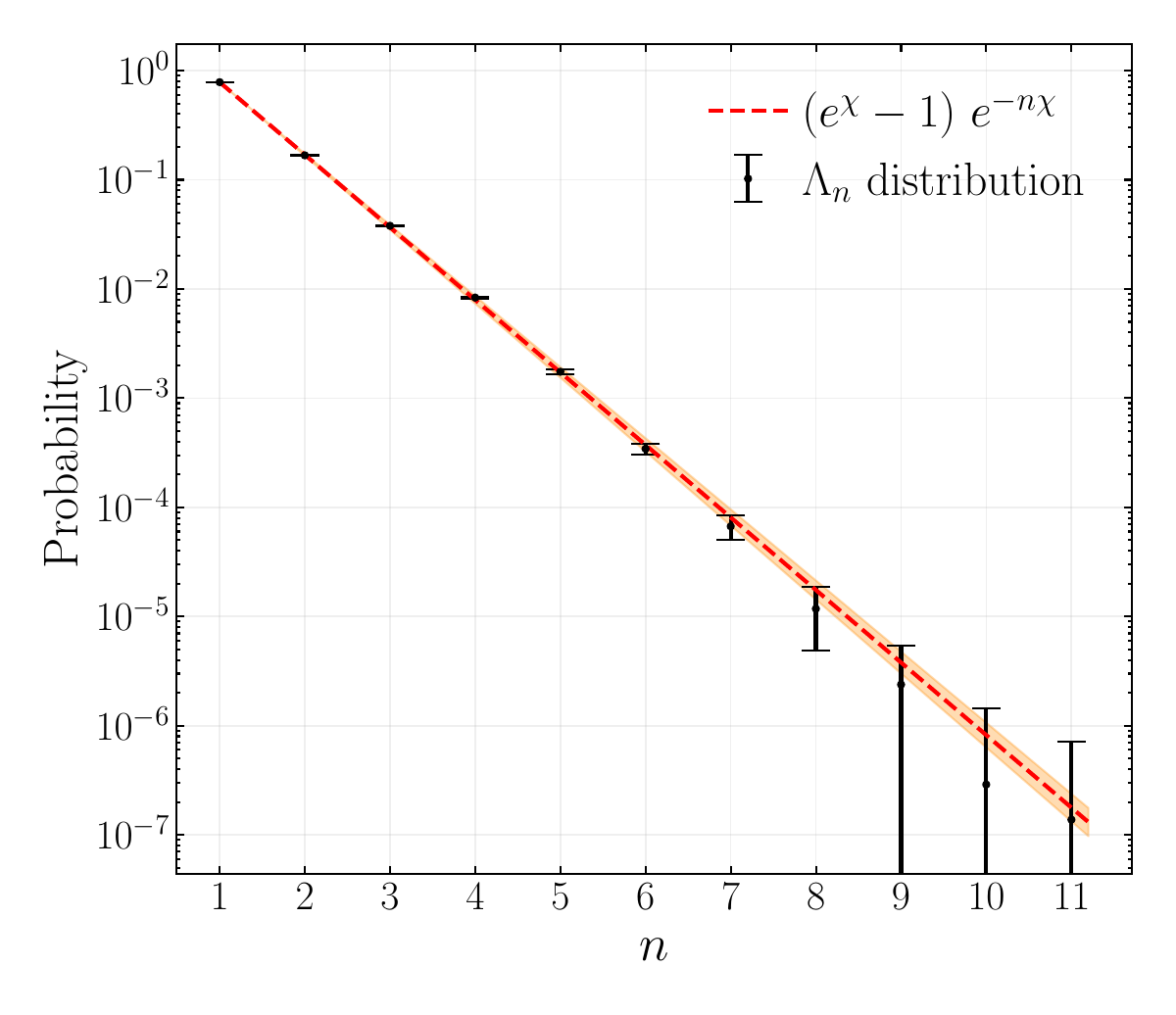}
        \put(-136, 88){(c)}
        \put(-2, 88){(d)}
    \end{overpic}
    \phantomsubfloat{\label{fig: lambda boundaries}}
    \phantomsubfloat{\label{fig: Nlambdas_vs_Area}}
    \phantomsubfloat{\label{fig: Nlinks_vs_area}}
    \phantomsubfloat{\label{fig: n_lambda_probability_distribution_expx_logy}}
    \vspace{-8mm}
    \caption{{ 
    Numerical results of our simulations.
    a) Averages of the maximal separation of elements forming horizon molecules from the relevant hypersurfaces, with $1\sigma$ and $3\sigma$ deviations highlighted. 
    Above, the maximum temporal separation $\Delta t_\mathrm{max}$ from $\Sigma : t^*=0$. 
    Below, the maximal spatial separation $\Delta r_\mathrm{max}$ between the horizon and the furthest elements in $\Lambda_n$ molecules, both from the inside and outside.
    b) Number of $\Lambda_n$ molecules $\langle N_{\Lambda_n}\rangle$ scaling linearly with the horizon area $A$. 
    The largest molecule's size we found was 11.
    c) Average umber of Links $\langle N_L \rangle$ (black) as a function of the horizon area $A$, and relative standard deviation (blue crosses).
    The azure and green dashed lines represent two analytical models: the analytical flat spacetime approximation from Eq.~(\ref{eq: avg expected nb of links 0order}), and
    the scaling with curvature correction from Eq.~(\ref{eq: Nlink corrected for curv}), respectively, both with $a^{0}_L = \sqrt{3}/10$.
    The red line is our best fit of the curvature-corrected expansion, specified in \cref{eq:best-fit}.
    Note, $A_{\ell} = A/\ell^2$.
    The first two models lie within the uncertainties but are consistently lower, as evident in the inset. 
    The red line is a perfect fit.
    d) The distribution of $\Lambda_n$ is well described by a falling exponential function $p_n\propto e^{-\chi n}$, where $\chi=1.53\pm0.03$.
    The yellow cloud around the fit shows the $1\sigma$ deviation.
    Likely due to the finiteness of our simulations, the number of molecules of size larger than 7, for which poor statistics was found, is slightly lower than the fit.}}
    \label{fig: allresults}
\end{figure*}

The $\Sigma$ hypersurface was set at $t^{*}=0$. 
We sprinkled into a (3+1)D hollow cylinder with points lying in $t^* \in [-4, 0]$ and $r\in[r_S - 3,\, r_S + 3]$ in terms of discreteness units $\ell$.
These bounds were chosen close enough to the $\Sigma$ and $\mathcal{H}$ hypersurfaces to allow for an efficient simulation of large black holes, but far enough to avoid affecting the behaviour of the molecules.
All points that are part of $\Lambda_n$ molecules consistently lie far from the chosen boundaries (see Fig. \ref{fig: lambda boundaries}).

The number of $\Lambda_n$ shows clear proportionality with the horizon area (see Fig. \ref{fig: Nlambdas_vs_Area}), giving the Pearson correlation coefficients $r\approx1$ for $n<8$, at which point we lack enough statistics. 
The number of Link molecules, which is a weighted sum of $\Lambda_n$, also scales linearly with $A$ (see Fig. \ref{fig: Nlinks_vs_area}).
Therefore, the proportionality constants we obtain are:
\begin{itemize}
    \item Neglecting curvature corrections, the proportionality coefficient is estimated as 
    $$a^{(0)}_{L[\text{1}]} = 0.1757 \pm 0.0004.$$
    \item Setting the curvature correction $a_L^{(1)}$ to the theoretical value $a^{(1)}_{L[\text{th}]}=-0.0558$, $a^{(0)}$  is estimated as $$a^{(0)}_{L[\text{2}]}=0.1761 \pm 0.0004$$.
    \item Fitting both $a^{(0)}$ and $a^{(1)}$ gives
    \begin{equation}
        a^{(0)}_{L[\text{3}]}=0.173 \pm 0.001, \qquad a^{(1)}_{L[\text{3}]}=0.4 \pm 0.2 \label{eq:best-fit}.
    \end{equation} 
\end{itemize}
Recall that the analytic result from Barton et al.~\cite{Barton2019} is $a^{(0)}_{L[\text{th}]}=\sqrt{3}/10 \approx 0.1732$. 
This perfectly agrees with the numerical fit when accounting for the 1st order curvature correction.
Oppositely, we note our best numerical estimate for $a^{(1)}_L$ differs from the analytic estimate both in order of magnitude and sign. 
This will be discussed later.
Finally, non trivially, \cref{fig: Nlinks_vs_area} also shows that the relative uncertainty in the number of Links $\sigma_L/\langle N_L\rangle$ is small and further decreases with the size of the black hole. 

We also studied the distribution of $\Lambda_n$ molecules (see Fig. \ref{fig: n_lambda_probability_distribution_expx_logy}).
Disregarding data for $n\geq8$ which has insufficient statistics due to the finiteness of our simulations, we discover that we can model the distribution very well by a falling exponential 
\begin{equation}\label{eq: nlambda distribution}
    p_n = \left(e^\chi - 1\right) e^{-n\chi}, \quad \chi = 1.53 \pm 0.03. 
\end{equation}
Using the relation between the number of Links and $\Lambda_n$ molecules and the aforementioned probability distribution, it is easy to infer that
\begin{equation}
    \langle N_\Lambda \rangle = \left(1- e^{-\chi} \right) \langle N_L \rangle.
\end{equation}
Inserting the numerical scaling factor $a^{(0)}_{L[\text{3}]} = 0.173 \pm 0.001$ into Eq.~(\ref{eq: Nlink corrected for curv}) and using the probability distribution of $\Lambda_n$ molecules, the Boltzmann entropy given by Eq.~(\ref{eq : Boltzmann entropy}) yields
\begin{equation}
    \label{eq: S_lambda}
    S_{\Lambda} = (0.091 \pm 0.002) \: \frac{A}{\ell^2}.
\end{equation}
Equating Eq.~(\ref{eq: S_lambda}) with the black hole entropy $S_{BH}$ from Eq.~(\ref{eq: BH formula}) we can give a physical meaning to the discreteness scale $\ell$. In terms of Plank length $\ell_p$, we obtain
\begin{equation}\label{eq: l=..l_p for lambdas}
    \ell = (0.603\pm0.003) \: \ell_p.
\end{equation}
Inverting, $G\hbar = (2.75\pm0.03) c^3 \ell^2$.

\section{Discussion}\label{Discussion}

This section discusses the obtained results and situates the paper into the landscape of black hole thermodynamics in Quantum Gravity.

\subsection{Numerical results}\label{subsec:results-discussion}

We confirmed that the entropy of the Schwarzschild black hole modelled by the horizon molecules unequivocally scales linearly with the horizon area for $A \gg \ell^2$, in agreement with the well-established Bekenstein-Hawking formula.
This is a very important result for two reasons. It numerically demonstrates for the first time that CST can yield correct predictions in curved (non-conformally flat) spacetime and that it can be used to study black hole thermodynamics.

The analytic estimate by Barton et al.~\cite{Barton2019} for $a_L^{0}$ perfectly matches our numerical fit. 
However, the analytic and numerical estimates for the 1\ts{st} order curvature correction disagree. 
The analytic model lies within our uncertainties but is systematically lower than our averages.
We note that a discrepancy was also found by Ref.~\cite{Narin} with the 1\ts{st} order correction due to horizon expansion in their study of a dynamical black hole in flat spacetime. 

Moreover, we were able to \lq\lq count'' the Planck length. 
If causal sets were the underlying structure from which the spacetime continuum emerges, and if correspondingly CST was the fundamental theory, then it would have no free parameters.
This would imply that Planck length $\ell_p$ is some multiple of the discreteness length $\ell$, which \textit{is} the fundamental length of size $1$ in the fundamental units. 
As generally anticipated, we found that $\ell \sim \ell_p$, which is where a lot of general physics breaks down. 

It is interesting to note the sizes of the black holes we simulated. 
The largest ones had an area of $A=42000 \: \ell^2$ and since $\ell\sim \ell_p$ this corresponds to mass $M = \sqrt{A/16\pi} \sim 30$ in Planck units, i.e. $\sim 10^{-7}$ kg. 
For comparison, the masses of the observed black holes usually range from $10^{30}$ kg to $10^{38}$ kg, yet, our tiny black holes yielded such good results.

Furthermore, we were intrigued by the fact that the falling exponential distribution for $\Lambda_n$ molecules had an exponent $\chi = 1.53 \pm 0.03$, very close to $3/2$ (see Fig. \ref{fig: n_lambda_probability_distribution_expx_logy}). 
If we treat $\Lambda_n$ molecules as thermal molecules that exist in $n$ different microstates,
we can write their thermal energy distribution as $p_n \propto e^{-E_n/k_B T}$.
Comparing this with Eq.~(\ref{eq: nlambda distribution}), we find that
\begin{equation}\label{eq: horizon molecule energy}
    E_n \approx \frac{3n}{2} k_B T,
\end{equation}
which is the energy of $n$-atomic molecules in 3D space.
Since $\Lambda_n$ molecules exist in 3D space at a given moment of time, we hoped to be able to generalise this expression, such that Eq.~(\ref{eq: horizon molecule energy}) would have been just a specific case for $f=3$ degrees of freedom.
However, simulations of causets in (2+1)D spacetime showed that $\chi \sim 1.5$ in (2+1)D as well (see \cref{app:other-numerical-results}). Therefore, it is false that $\chi \sim f/2$ in all dimensions, and $\Lambda_n$ molecules cannot be interpreted as thermal molecules.

\subsection{Holographic principle and information paradox}

Hawking argued that black hole radiation only depends on its external degrees of freedom~\cite{Hawking1976}: mass, charge and angular momentum. 
By emitting radiation, black holes can evaporate away.   
In this case, all that is left, the radiation, thus retains only information about those three features. All other information about previously absorbed objects is lost.
This breaks the fundamental law of conservation of quantum information~\cite{Hawking1976, Banks_1984}, and constitutes the black hole information paradox.

There are two possible solutions: either black hole evaporation stops at the Planck scale, at which Hawking's derivation breaks down, or radiation contains more information than assumed. 
Recently, different suggestions~\cite{Lowe_1999, Hawking2005, Hawking2015Information, Hawking2016SoftHair} have been made in support of the latter.
These follow, albeit in different ways, from the holographic principle, which states that a complete description of a volume of space is given by the degrees of freedom of its surface~\cite{Hooft1993, Susskind1995}. 
Horizon molecules are distributed over the horizon's surface, within an infinitesimal shell of thickness $\sim 4 \ell$.
They \textit{form} the horizon.
As they encode the entropy of the black hole, they might provide a relation to the holographic principle and the information paradox. 
This consideration requires further studies.

\subsection{Planckian black holes}
\label{sec:PlanckianBH}
Black holes form from the gravitational collapse of massive stars. 
They form with a mass of at least the order of the mass of the Sun~\cite{BHSize_Thompson2019}, hence at most a Hawking temperature of order $\sim 10^{-8}$K. 
This is 100 million times colder than the Cosmic Microwave Background (CMB)~\cite{Fixsen2009}. 
Therefore, the energy black holes absorb from CMB surpasses the losses from their own radiation, making it impossible for current black holes to evaporate into smaller masses. 

However, the CMB will drop. Therefore, black holes will eventually be allowed to evaporate. 
Furthermore, it was suggested that primordial black holes of Planckian size could have formed in the initial stages of the universe via different processes~\cite{Hawking1971_lowmass}. 
Hawking's work breaks down at Planckian scales, begging the question of how a black hole of Planckian size behaves. 
In this section, we provide an answer within the theory of horizon molecules.
Given that the molecular model might break down at such small scales, especially given the current lack of quantum effects in the picture, the results of this subsection need to be taken \textit{cum grano salis}. 
Nevertheless, we believe they might offer interesting insights for further research.

To answer in the context of our molecular approach to black hole entropy, consider that, in general, the average number of horizon molecules is a sum of the form
\begin{equation}
    \langle N \rangle = 
    \sum_{i=0}^{\infty} a^{(i)} \left(\frac{\sqrt{A}}{\ell}\right)^{2-i}, \label{Eq.aLinks2}
\end{equation}
where $(i)$ labels the order of the curvature correction - with $i=0$ being the flat spacetime approximation and $a^{(i)}$ a real constant for all $i$. 
This implies an entropy
\begin{equation}
    S = 
    \sum_{i=0}^{\infty} \sigma^{(i)} \left(\frac{\sqrt{A}}{\ell}\right)^{2-i}, \label{eq:Entropy_with_sigmas_i}
\end{equation}
where, if the molecules have distribution $\{p_n\}$, it yields that $\sigma^{(i)} = - a^{(i)} \sum_{n} p_n \ln{p_n}$.
The leading term gives the standard black hole entropy, hence $\ell^2=4\sigma^{(0)}$ and 
\begin{equation}
    S = 
    \frac{A}{4} 
    + \sum_{i=1}^{\infty} \sigma^{(i)} \left(\frac{\sqrt{A}}{\ell}\right)^{2-i}.
\end{equation}
Thus, for a Schwarzschild black hole $A=16\pi M^2$ we can write
\begin{equation}
    S = 
    4 \pi M^2 
    + \sum_{i=1}^{\infty} 
    \varsigma^{(i)} \left(\frac{M}{\ell}\right)^{2-i}, \label{Eq.entropySch}
\end{equation}
where $\varsigma^{(i)} = (4\sqrt{\pi})^{2-i} \sigma^{(i)}$. Applying the mass-energy equivalence $U=M$, the 1\ts{st} law of thermodynamics gives
\begin{equation}
    \frac{1}{T} = \frac{\partial S}{\partial M}. \label{eq:1_over_T_dS_dM}
\end{equation}
Hence, by rearranging we obtain
\begin{equation}
    \frac{1}{T} = 
    8\pi M + \frac{\varsigma^{(1)}}{\ell} - 
    \frac{1}{\ell}\sum_{i=3}^{\infty} 
    (i-2) \varsigma^{(i)} \left(\frac{\ell}{M}\right)^{i-1}. \label{Eq.aTemp}
\end{equation}

The first term dominates for large $M \gg \ell$, in which case this results in the Hawking temperature. 
The second represents a correction of order unity. In the limit $M \rightarrow 0$, the second term alone could give a cutoff at approximately the Planck temperature $T_p \sim 10^{32} K$. A similar cutoff is predicted by loop quantum gravity too, albeit due to different mechanisms~\cite{Röken_2013}.

Before investigating the effects of the third contribution, let's outline the constraints on the real constants $\varsigma^{(i)}$.
Define ${\mu} = \ell/M \in (0, \mu_d]$, where $\mu_d$ could be infinite as $M$ vanishes. 
We require $\forall \mu \in (0, \mu_d]$:
\begin{enumerate}[label=(\roman*)]
    \item Entropy Positiveness. 
    From Eq.~(\ref{Eq.entropySch}), relabelling the summation index with $j = i-2$ and taking out the first two elements, 
    \begin{equation}
        \quad S = \frac{4 \pi \ell^2}{\mu^2} + \frac{\varsigma^{(1)}}{\mu} + \varsigma^{(2)} + \sum_{j=1}^{\infty} \varsigma^{(j+2)} \mu^{j} \; \geq 0. \label{eq:entropy_varsigma_mu}
    \end{equation}

    \item Entropy Finiteness. 
    From Eq.~(\ref{eq:entropy_varsigma_mu}), 
    \begin{equation}
        \sum_{j=1}^{\infty} \varsigma^{(j+2)} \mu^{j} \;\; \mathrm{is\;\;finite\;}.
    \end{equation}

    \item Temperature Positiveness. 
    From Eq.~(\ref{Eq.aTemp}),
    \begin{equation}
        \frac{1}{T} = \frac{8\pi \ell}{\mu} + \frac{\varsigma^{(1)}}{\ell} - \frac{\mu}{\ell} \; \sum_{j=1}^{\infty} j \; \varsigma^{(j+2)} \mu^{j} \; \geq 0. \label{eq:T_varsigma_mu}
    \end{equation}
\end{enumerate}
A type of molecule could perhaps be envisaged such that black holes are not allowed to fully evaporate. This would be possible if the following condition is also satisfied:
\begin{enumerate}[label=(\roman*)]
    \setcounter{enumi}{3}
    \item Temperature Loss. 
    From Eq.~(\ref{eq:T_varsigma_mu}), $T\rightarrow0^{+}$ for a non-zero mass if there exists a finite $\mu_d$ such that
    \begin{equation}
        \sum_{j=1}^{\infty} j \; \varsigma^{(j+2)} \mu_d^{j} \;
        \rightarrow -\infty  \; 
    \label{Eq.TempVanish}
    \end{equation}
\end{enumerate}
Horizon molecules with real constants $\varsigma^{(i)}$ satisfying conditions (i-iv) would give an always positive finite entropy and a temperature which is always positive and finite but for $\mu = \mu_d$, at which it vanishes. Physically, the black hole would radiate less and less, asymptotically approaching a minimum mass $M_{\mathrm{min}} = \ell / \mu_d$, thus preventing full evaporation. The black hole would reach an equilibrium, at a finite temperature, when the emission matches the ingoing radiation from the CMB: the black hole would be a tiny dark fragment wandering in the universe. 

This could resolve the information paradox. The information related to the absorbed material could be stored inside the horizon, forever. Furthermore, as black holes can form where none was previously present, a mechanism preventing their complete evaporation could break charge-parity-time-reversal (CPT) symmetry~\cite{Hawking1976}. 

Can such a set of real constants $\varsigma^{(i)}$ exist? 
It can. 
As an example, consider the Riemann zeta function
\begin{equation}
    \zeta(x) = \sum_{i=1}^{\infty} i^{-x}.
\end{equation}
Assume that $\varsigma^{(i+2)} = ai^{-x} \;\;\forall i\in \mathbf{N}$, where $a$ and $x$ are real constants. This leaves $\varsigma^{(1)}$ and $\varsigma^{(2)}$ momentarily undefined. 

Then, the sums in (ii) and (iv) are $a\mathrm{Li}_{x}(\mu)$ and $a \mathrm{Li}_{x-1}(\mu_d)$ respectively, where
\begin{equation}
    \mathrm{Li}_{x}(\mu) =  \sum_{j=1}^{\infty} j^{-x} \mu^{j}
\end{equation}
is the polylogarithm~\cite{Lewin1981}. $\mathrm{Li}_{x}(\mu)$, for real $x$ and $\mu$, converges for $\mu<1$, or for $\mu=1 \land x > 1$. Then (ii) converges for $\mu_d<1 \; \forall x$ or for $\mu_d = 1 \land x > 1$. 
(iv) diverges for $\mu_d > 1 \; \forall x$ or for $\mu_d = 1 \land x \leq 2$. 
For it to be negative, $a<0$ is needed. Therefore, (ii) and (iv) are both satisfied for $\mu_d = 1, \; x\in(1,2]$ and $a<0$. 

These satisfy (i) and (iii) as well. 
Taking $\varsigma^{(1)}$ and $\varsigma^{(2)}$ to be positive, the only negative contribution in (i) comes from the series, which converges since (ii) is satisfied. With $|a|$ small enough, this negative term is smaller than others, hence (i) is satisfied. Finally, as $a<0$ and $\mu>0$, the series in (iii) is necessarily positive, hence (iii) is satisfied.

Therefore, it is possible to find real constants $\varsigma^{(i)}$ satisfying conditions (i-iv). 
Whether a molecule with such constants exists remains an open question. 
It must also be noted that, at Planckian scales, the average number of molecules is small, 
thus, the validity of Eq.~(\ref{eq:Entropy_with_sigmas_i}) in this limit is questionable in the first place. 
Moreover, as noted at the beginning of the subsection, the molecular model itself is vague at such small scales.
Nevertheless, this result hints at the potential of CST to express new physics and provides insights for further research.
As computing curvature corrections of arbitrary order seems an overwhelmingly hard task, numerical studies on the behaviour of $\partial S/\partial M$ in the small $M$ limit are suggested.

\section{Conclusion}\label{Conclusion}

This paper shows for the first time a numerical study of the entropy of a Schwarzschild black hole in the framework of Causal Set Theory, an approach to quantum gravity which only adopts the assumptions of
\begin{enumerate}[label=(\Roman*)]
    \item spacetime discreteness,
    \item transitive and acyclic causal relations,
    \item Poissonian correspondence between discrete and continuum.  
\end{enumerate}

We simulated causal sets in Schwarzschild spacetime following the procedure outlined in section \ref{sec: causality in 3+1 D spacetime}, which corrects the algorithms provided by the original paper~\cite{He2009}. 
In order to efficiently simulate causal sets we created a highly parallelized computational framework in \texttt{C++}, described in \cref{Simulation Framework}, which allowed us to generate causets in non-conformally flat Schwarzschild spacetime orders of magnitude larger than any previous simulation. 

We showed for the first time that the minimal assumptions of Causal Set Theory are enough to provide a molecular model of black hole entropy where the horizon molecules
\begin{enumerate}[label=(\roman*)]
    \item are localised close to the horizon within $2\ell$ distance,
    \item give a black hole entropy proportional to the horizon area in the $A\gg\ell^2$ limit which agrees with the Bekenstein-Hawking formula,
    \item imply a discreteness length $\ell$ of Planckian order $l_p \sim 10^{-35}$m. 
\end{enumerate}

Furthermore, we showed that the numerical scaling of Link molecules agrees perfectly with the analytical result obtained by Barton et al.~\cite{Barton2019} in the flat spacetime approximation.
Oppositely, the 1\ts{st} order curvature corrections, which we analytically computed in section \ref{sec: Horizon Molecules}, were found to disagree with our numerical estimate. 
Specifically, the analytical result lies within the uncertainties of data points, but it is systematically lower. 
However, we note this discrepancy does not impact the aforementioned conclusions of our study, especially for black holes of physically relevant sizes, where the curvature correction is negligible.

Lastly, we argued that the horizon molecule approach may yield a finite temperature cut-off, and even prevent full black hole evaporation for black holes of Planckian size. 
Whether a horizon molecule can be envisaged with these properties remains an open challenge. 
Moreover, we warn that these results need to be taken with caution, as the validity of the considered molecular model is likely to break down at sizes close to the Planck scale, especially since quantum effects are yet to be introduced.
Nevertheless, these conclusions are intriguing and may offer valuable insights.

There still lies a myriad of exciting unanswered questions about black hole thermodynamics and causal sets in non-conformally flat spacetimes. 
Given the simulation framework now publicly available on \href{https://github.com/vidh2000/MSci_Schwarzschild_Causets}{\texttt{GitHub}}, we hope to have opened the door for future causal set studies in curved spacetime, possibly discretising more complex black holes, such as Riessner-Nordstrom and Kerr, investigating other entropy models~\cite{BenincasaPhD, Sorkin2018}, and exploring other curvature features~\cite{Benincasa2010, Roy2013}.

\begin{acknowledgements}
We would like to thank Prof Fay Dowker, who developed the idea of the Boltzmannian state counting in causal sets theory, for her guidance, the fun discussions, and the extensive feedback on the manuscript. 
We would also like to thank Hinchi Lee, Narin Tananitaporn, Rohith Mukkapati and Olivia Ho for motivating us and whose ideas contributed to the success of our project.
\end{acknowledgements}
%
\appendix

\section{Considerations on causality in (3+1)D}\label{app:4Dcausality}

In this appendix, we give mathematical proof of the final statements of Section \ref{sec: causality in 3+1 D spacetime}. There we concluded that we can fit 
\begin{equation}
    \label{eq:app:dphidu}
    \frac{d\phi}{du} = \pm_{du} \left(\eta^2 + 2Mu^3 - u^2 \right)^{-1/2}
\end{equation}
against the spatial coordinates of the events to obtain $\pm \eta$, and then use
\begin{equation}
    \frac{dt^*}{du} = \frac{1}{u^2(2Mu-1)} 
    \left[
    \frac{\pm_{du}\eta}{\sqrt{\eta^2 + u^2(2Mu-1)} } + 2Mu \right]
    \label{eq:app:dt*du}
\end{equation}
to find time bounds on causality, where
\begin{equation}
    \pm_{du} = \mathrm{sign}(u_2 - u_1).
\end{equation}
Making the signs of $\eta$ and $dr$ explicit, and re-converting the above equations in terms of $r=1/u$,
we have
\begin{equation}
    d\phi
    = \frac{\pm_{dr}}{r^{2}}  \left[\eta^{2} - \left(1-\frac{2M}{r}\right)r^{-2}\right]^{-1/2} dr,
    \label{eq:app:dphidr}
\end{equation}
and
\begin{equation}
    dt^*=\frac{|dr|}{\frac{2M}{r}-1} \cdot
    \begin{cases}
        \frac{|\eta|}{\sqrt{\eta^{2} + (\frac{2M}{r} - 1)r^{-2}}}
        - (\pm_{dr} \frac{2M}{r}),
        & \text{$\eta > 0$}
        \\ \\
        \frac{-|\eta|}{\sqrt{\eta^{2} + (\frac{2M}{r} - 1)r^{-2}}}
        - (\pm_{dr} \frac{2M}{r}),
        & \text{$\eta < 0$}
    \end{cases} 
    \label{eq:dt*dr2}
\end{equation}

Next, we show which bounds are to be considered based on where the events lie with respect to the horizon, and show that when fitting Eq.~(\ref{eq:app:dphidu}) we require $\phi \in [0, 2\pi)$.

\subsection{Both points inside the horizon}

When \textit{both} points are \textit{inside} the horizon, $2M>r$. As expected, our equations force any object inside the horizon to always keep falling towards the singularity.
In fact, consider a movement opposite to the fall, with $r_2 > r_1$, hence $dr > 0$. We impose $dt^*>0$, as $t_2^* > t_1^*$. From Eq.~(\ref{eq:dt*dr2}), as $2M>r$, this requires
\begin{equation}
    \frac{\eta}{ \sqrt{ \eta^{2} + (2M/r - 1)r^{-2} } } > \frac{2M}{r}.
\end{equation}
Trivially, this has no solution for $\eta<0$. For positive $\eta = |\eta|$, substituting $2M/r = 1 + \delta$, with $\delta>0$, and rearranging, this yields
\begin{equation}
    |\eta| > (1+\delta) |\eta| \sqrt{1 + \frac{\delta}{\eta^{2}r^2}},
\end{equation}
which also has no solution: objects inside the horizon keep falling in. 

Therefore, we must have $dr<0$, hence $\mathrm{sign}(r_2-r_1)<0$. Then, Eq.~(\ref{eq:dt*dr2}) shows that the negative and positive values of $\eta$ give, respectively, the smallest and largest times $t_-^*$ and $t_+^*$ of intersection with the spatial coordinates of $E_2$. Therefore, the events are related if $t_-^* \leq t_2^* \leq t_+^*$, similarly to the 2D case. He and Rideout~\cite{He2009} do not mention that both signs have to be considered, again neglecting the upper bound.

\subsection{One or both points outside the horizon}

If \textit{both} points are \textit{outside} the horizon, substituting $2M/r=\epsilon<1$ in Eq.~(\ref{eq:dt*dr2})
\begin{equation}
    dt^*=\frac{|dr|}{1-\epsilon} \cdot 
    \begin{cases}
        \frac{- |\eta|}{\sqrt{\eta^{2} - (1 - \epsilon)r^{-2}}}
        + \mathrm{sign}(r_2 - r_1) \epsilon
         & \text{if $\eta > 0$}
        \\ \\
        \frac{+|\eta|}{\sqrt{\eta^{2} - (1 - \epsilon)r^{-2}}}
        + \mathrm{sign}(r_2 - r_1) \epsilon
        & \text{if $\eta < 0$}
    \end{cases}. 
    \label{eq:dt^*dr_out}
\end{equation} 
As $\epsilon<1$, we see that
\begin{equation}
    \frac{|\eta|}{\sqrt{\eta^{2}-(1-\epsilon)r^{-2}}}>1>\epsilon.
\end{equation}
Then, $\eta>0$ gives a negative $dt^*$ for both values of $\mathrm{sign}(r_2 - r_1)$. This is useless for us, as we impose $t_2^*>t_1^*$ i.e. $dt^*>0$. Therefore, we use the negative $\eta$. This yields the earliest geodesic reaching $E_2$'s spatial coordinates. Thus, $E_1$ and $E_2$ are connected if $t_2^*>t^*$, where $t^*$ is the solution to the differential equation. 

As in 2D, this result also applies to the case when \textit{only} $E_1$ is \textit{outside} the horizon, because outside the horizon an object can remain at rest for an arbitrarily long period of time, hence no upper bound applies.

\subsection{The Optimal \texorpdfstring{$\phi$}{phi}}

Null geodesics might travel an angular distance $\phi_2+2k\pi$, matching the spatial coordinates of the events by wrapping around the singularity $k$ times. Fitting Eq.~(\ref{eq:app:dphidr}), a set $\{\eta^{2}_k\}$ of suitable parameters exists, one for each number $k$ of wrappings. We show that the best $\eta^{2}$ corresponds to the 0 wrappings case, i.e. fitting Eq.~(\ref{eq:app:dphidr}) to $\phi_2$.

First note that with increasing $k$, as $|\Delta r|$ stays the same whereas $|\Delta \phi|$ increases, a larger $|d\phi/dr|$ is required. This implies, from Eq.~(\ref{eq:app:dphidr}), $|\eta_k|$ decreases with $k$.
Then, let's write Eq.~(\ref{eq:dt*dr2}) as
\begin{equation}
    dt^* = \left[\;a^{-1}(r) \mathrm{sign}(\eta) f\left(|\eta|; r\right) + b(r) \; \right] |dr|, 
\end{equation}
where $a(r) = (2M/r - 1)$ and
\begin{equation}
    f\left(|\eta|; r\right) = \frac{|\eta|}{\sqrt{\eta^{2} + r^{-2}a(r)}} 
    = \frac{1}{\sqrt{1+r^{-2}a(r)/\eta^{2}}}.
\end{equation}
For both events inside the horizon, $a(r)>0$, hence $f$ monotonically increases with $|\eta|$, and so does $a(r) f$. Then the largest $|\eta|$ yields the earliest ($\eta<0$) and latest ($\eta>0$) null geodesics, thus setting the boundaries of the light cone. This is given by $k=0$ wrappings.

For both events outside the horizon, $a(r)<0$, hence $f$ monotonically decreases with $|\eta|$, hence $a(r) f$ monotonically increases. Thus, the largest $|\eta|$ possible -corresponding to 0 wrappings, again sets the earliest geodesic. Recall, outside the horizon $\eta<0$ is the only relevant solution. 

\section{Curvature corrections in Schwarzschild spacetime}\label{appendix: curv corrections}

In \cref{sect: nlambda mulecules}, we derived the analytic scaling for the number of Link molecules with horizon area in the flat spacetime approximation.
In this appendix, we derive the 1\ts{st} order curvature correction.
Therefore, Eq.~(\ref{eq: avg expected nb of links 0order}) is the 0$^{\text{th}}$ order approximation to the full expression in Schwarzschild spacetime given by

\begin{equation}\label{eq: Nlink expanded}
 \langle N_{L}^{BH}\rangle = a_{L} \frac{A}{\ell^2} = \left(a_{L}^{(0)}+a_{L}^{(1)}\frac{\ell}{\sqrt{A}} + \mathcal{O}\left(\frac{\ell^2}{A}\right)\right) \frac{A}{\ell^2}.  
\end{equation}
Barton et. al~\cite{Barton2019} outlined a general procedure to derive the 1$^\text{st}$ order curvature correction term $a_L^{(1)}$ for arbitrary horizon.
Here we derive it for the Schwarzschild spacetime. 
The final result of this derivation was shown in Eq.~\ref{eq: Nlink corrected for curv}.

\subsection{Preliminaries}

The horizon $\mathcal{H}$ of the Schwarzschild black hole and the spacelike hypersurface $\Sigma$ are in Eddington-Finkelstein-original coordinates defined as
\begin{equation}
    \mathcal{H}(x^\mu): r = r_S = 2M, \quad \quad \Sigma(x^\mu): t^* = 0.
\end{equation}
We denote the intersection of the two hypersurfaces as $\mathcal{J}$ and it represents the event horizon at time $t^*=0$. Hence, evaluating tensors at $\mathcal{J}$ implies setting $r=r_S=2M$ and $t^*=0$.
Moreover, for completeness the metric $g_{\mu\nu}$ and its inverse $g^{\mu\nu}$ in EFO coordinates are
\begin{equation}\label{eq: metric}
    g_{\mu\nu}=
    \begin{pmatrix}
        -\left(1-\frac{2M}{r}\right) & \frac{2M}{r} & 0 & 0\\
        \frac{2M}{r} & \left(1+\frac{2M}{r}\right) & 0 & 0 \\
        0 &  0 & {r^2} & 0 \\
        0 & 0 & 0 & {r^2 \sin^2{\theta}}
    \end{pmatrix},
\end{equation}
and
\begin{equation}
    g^{\mu\nu}=
    \begin{pmatrix}
        -\left(1+\frac{2M}{r}\right) & \frac{2M}{r} & 0 & 0\\
        \frac{2M}{r} & \left(1-\frac{2M}{r}\right) & 0 & 0 \\
        0 &  0 & \frac{1}{r^2} & 0 \\
        0 & 0 & 0 & \frac{1}{r^2 \sin^2{\theta}}
    \end{pmatrix}.
\end{equation}
The 1$^\text{st}$ order corrections must depend on mutually independent invariant scalars of dimension $\ell^{-1}$, evaluated at $\mathcal{J}$ and depend on the geometry of the problem at hand.
The only such quantities are $K$, the trace of the extrinsic curvature of $\Sigma$, its component $K_m$ tangential to $\Sigma$ an orthogonal to $\mathcal{J}$, and the null expansion $\vartheta$ of the horizon. It can be shown that the 1$^\text{st}$ order correction for links in 3+1 dimension is~\cite{Barton2019}
\begin{equation}
    \label{eq: curv scale correction expression inappendix}
    \frac{a_L^{(1)}}{\sqrt{A}} \approx -  (0.036 K + 0.088 K_m + 0.021\vartheta).
\end{equation}
\vspace{-1mm}

\subsection{Determining the required geometrical objects}

To determine the invariant scalars we first require finding the basic geometrical objects from which they are constructed.
The first one is the metric, given by Eq.~(\ref{eq: metric}).

Then we also need $n^\mu$ and $k^\mu$, normal vectors to the hypersurfaces $\Sigma$ and $\mathcal{H}$, respectively. 
Vector $n^\mu$ is future-pointing ($n^0>0$), normalised as $n^\mu n_\mu =-1$ and is given by
\begin{equation}
    n_\mu = a\: \partial_\mu \Sigma = a (1,0,0,0),
\end{equation}
Correspondingly, we have 
\begin{equation}
    n^\mu = g^{\mu\nu} n_\nu = a \left(-1-\frac{2M}{r}, \frac{2M}{r},0,0\right).
\end{equation}
The previous conditions yield 
\begin{equation}
    a = \frac{-1}{\sqrt{\frac{2M}{r} + 1}}.
\end{equation}
Similarly, $k^\mu$ is also future-pointing and is given by
\begin{equation}
    k_\mu = b\:\partial_\mu \mathcal{H} = b(0, 1, 0, 0).
\end{equation}
Since it is normalised such that $[n^\mu k_\mu]_{\mathcal{J}} = -1/\sqrt{2}$, where $[\,\cdot\,]_{\mathcal{J}}$ denotes that the quantity is evaluated at $\mathcal{J}$, this immediately sets $b=1$.

Lastly, we require the vector $m^\mu = \sqrt{2}k^\mu - n^\mu$, which is tangent to $\Sigma$ and orthogonal to $\mathcal{J}$.

\subsection{Local geometric invariants}

Next, we introduce the \lq\lq projector'' tensors 
\begin{equation}
    h^\alpha_{\;\;\beta} = \delta^\alpha_\beta + n^\alpha n_\beta
\end{equation}
and
\begin{equation}
    \sigma^\alpha_{\;\;\beta} =  h^\alpha_{\;\;\beta} - m^\alpha m_\beta.
\end{equation}
Noting that extrinsic curvature is defined as
\begin{equation}
    K_{\mu\nu} = (\nabla_\beta n_\alpha) 
                h^\alpha_{\;\;\mu}
                h^\beta_{\;\;\nu},
\end{equation}
we can therefore determine the mutually independent invariant scalars evaluated on $\mathcal{J}$ that are required to obtain the curvature correction. 
The relevant Christoffel symbols needed for the computation are given in~\cref{CB_Chris}.
The trace $K$ of the extrinsic curvature is given by
\begin{equation}
    K(\mathcal{J}) = [g^{\mu\nu} K_{\mu\nu}]_\mathcal{J},
\end{equation}
its component along the $m$ direction by
\begin{equation}
    K_{m}(\mathcal{J}) = [K_{\mu\nu}m^\mu m^\nu]_\mathcal{J},
\end{equation}
and the null expansion of the horizon as
\begin{equation}
    \vartheta= [(\nabla_\beta k_{\alpha}) \sigma^{\alpha\beta}]_\mathcal{J}.
\end{equation} 
After determining the required Christoffel symbols in EFO coordinates and other required calculations, we obtain that $\vartheta = 0$, as expected since the Schwarzschild black hole is static.
The extrinsic curvature is found to be
\begin{equation}
    [K_{\mu\nu}]_{\mathcal{J}} = \frac{1}{32}
    \begin{pmatrix}
        \frac{3\sqrt{2}}{M} & \frac{6\sqrt{2}}{M} & 0 & 0\\
        \frac{6\sqrt{2}}{M} & \frac{12\sqrt{2}}{M} & 0 & 0\\
        0 & 0 & -32\sqrt{2}M & 0 \\
        0 & 0 & 0 & -32\sqrt{2}M \sin^2{\theta}
    \end{pmatrix},
\end{equation}
its trace is given by
\begin{equation}
    K = - \frac{5\sqrt{2}}{16} \frac{1}{M} \approx - \frac{0.4419}{M},
\end{equation}
and the component along $m^\mu$ is
\begin{equation}
    K_m = \frac{3\sqrt{2}}{16} \frac{1}{M} \approx \frac{0.2651}{M}.
\end{equation}
Inserting the obtained $K$, $K_m$ and $\vartheta$ into Eq.~(\ref{eq: curv scale correction expression}) and noting that $\sqrt{A}=4 \sqrt{\pi} M$ yields the 1$^\text{st}$ order curvature correction term
\begin{equation}
    \label{eq: link curv correction inappendix}
     a_L^{(1)} \approx  -0.0558.
\end{equation}
Then, the average expected number of Links in Schwarzschild spacetime, as a function of the horizon area $A$ in terms of the discreteness scale $\ell$, is given by
\begin{equation}
     \label{eq: Nlink corrected for curv appendix result}
    \langle N_{L}^{BH}\rangle = a_{L}\frac{A}{\ell^2} = 
    a_L^{(0)}\frac{A}{\ell^2} \left( 1 -0.322 \frac{\ell}{\sqrt{A}} + \mathcal{O}(\ell^2) \right),
\end{equation}
where $a_L^{(0)}=\sqrt{3}/10$ is the $0^{\text{th}}$ order flat spacetime term.
The correction term becomes negligible when $A$ increases to the size of ordinary black holes since $\ell\sim \ell_p$ is infinitesimal.

\begin{figure*}                
    \includegraphics[width=0.325\textwidth]{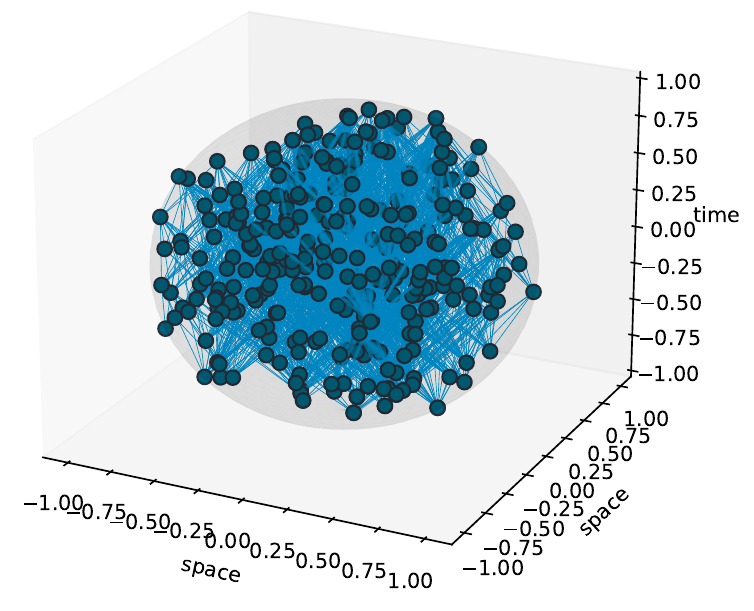}
    \includegraphics[width=0.325\textwidth]{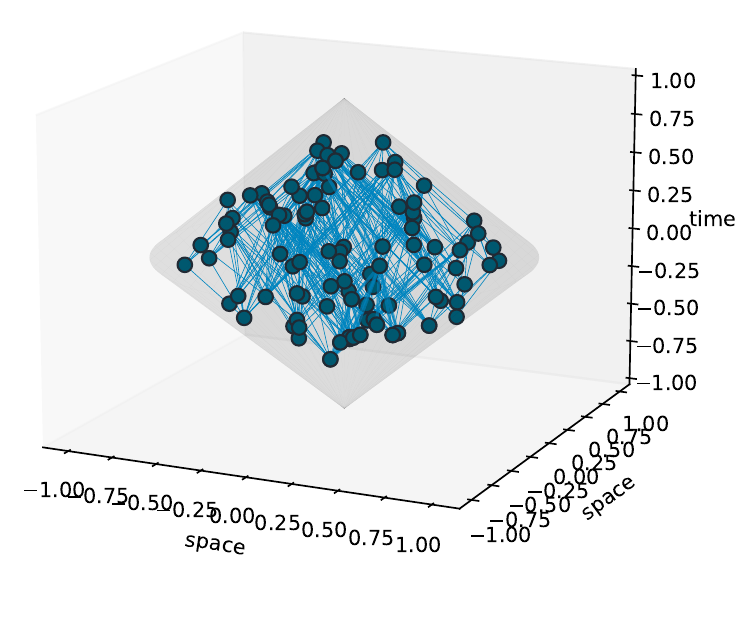}    
    \includegraphics[width=0.325\textwidth]{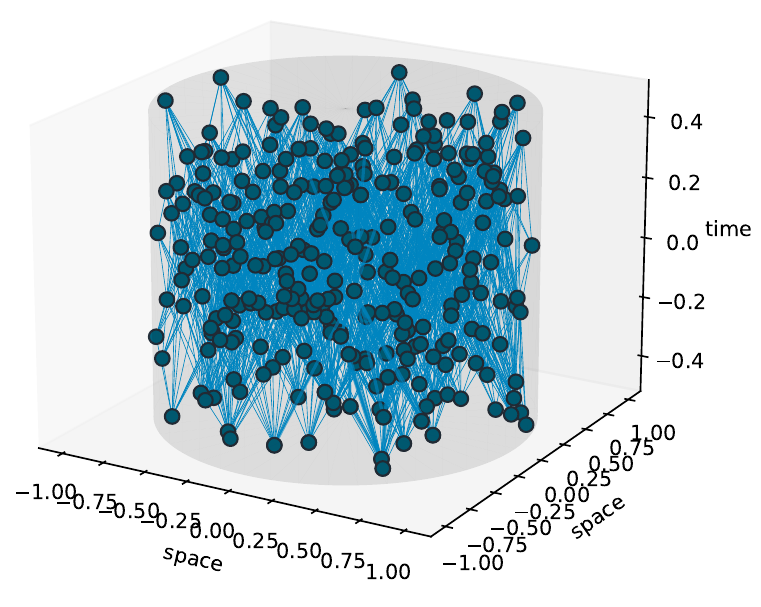}

    \phantomsubfloat{\label{fig: ball sprinkled}}
    \phantomsubfloat{\label{fig: bicone sprinkled}}
    \phantomsubfloat{\label{fig: cylinder sprinkled}}
    \vspace{-7mm}
    \caption{{Sprinkled causet elements in (2+1)D Minkowski spacetime lie within the shaded spacetime regions. The simulation framework allows for sprinkling into ball, bicone and cylinder shapes (left to right) in $D\in \{2,3,4\}$ spacetime.}}
    \label{fig: sprinkling test}
\end{figure*}

\section{Christoffel symbols in the Eddington-Finkelstein Original coordinates} \label{CB_Chris}

We computed the Christoffel symbols of the EFO coordinates introduced in Section \ref{sec:schwarz-spacetime} 
using \texttt{SymPy} and \texttt{EinsteinPy} libraries~\cite{SymPy, EinsteinPy}. 
The non-zero ones are
\begin{align}
    \Gamma^{t^*}_{{t^*}{t^*}} &= \frac{2M^2}{r^3}, \nonumber\\
    \Gamma^{t^*}_{{t^*}r}     = \Gamma^{t^*}_{r{t^*}} &= \frac{M(2M+r)}{r^3}, \nonumber\\
    \Gamma^{t^*}_{rr}         &= \frac{2M(M+r)}{r^3}, \nonumber\\
    \Gamma^{t^*}_{\theta \theta} &= -2M, \nonumber\\
    \Gamma^{t^*}_{\phi \phi} &= -2M\; \sin^2{\theta}, \nonumber
\end{align}
\begin{align}
    \Gamma^r_{{t^*}{t^*}} &= \frac{-M(2M-r)}{r^3}, \nonumber\\
    \Gamma^r_{{t^*}r}     = \Gamma^r_{r{t^*}} &= - \frac{2M^2}{r^3}, \nonumber\\
    \Gamma^r_{rr}         &= \frac{-M(2M+r)}{r^3}, \nonumber\\
    \Gamma^r_{\theta \theta} &= 2M-r, \nonumber\\
    \Gamma^r_{\phi \phi} &= (2M-r)\; \sin^2{\theta}, \nonumber
\end{align}
\begin{align}
    \Gamma^\theta_{r\theta} = \Gamma^\theta_{\theta r} &= \frac{1}{r}, \nonumber\\
    \Gamma^\theta_{\phi \phi} &= -\sin{\theta} \cos{\theta}, \nonumber
\end{align}
\begin{align}
    \Gamma^\phi_{r\phi} = \Gamma^\phi_{\phi r} &= \frac{1}{r}, \nonumber\\
    \Gamma^\phi_{\theta \phi} = \Gamma^\phi_{\phi \theta} &= \frac{1}{\tan{\theta}}. \nonumber
\end{align}

\section{Other tests of the causal set simulation framework}\label{appendix: Extra simulation framework tests}

This appendix describes other more basic and less relevant tests we performed, besides the ones from \cref{Simulation Framework}, to ensure our simulation framework works as intended for studying Schwarzschild causets.
 
Plotting the sprinkling region as a shaded shape, we confirmed that sprinkling into different shapes works as intended and that causets only lie within the required boundaries (see Fig. \ref{fig: sprinkling test}). 

\begin{figure}[h]
    \centering
    \includegraphics[width=\linewidth]{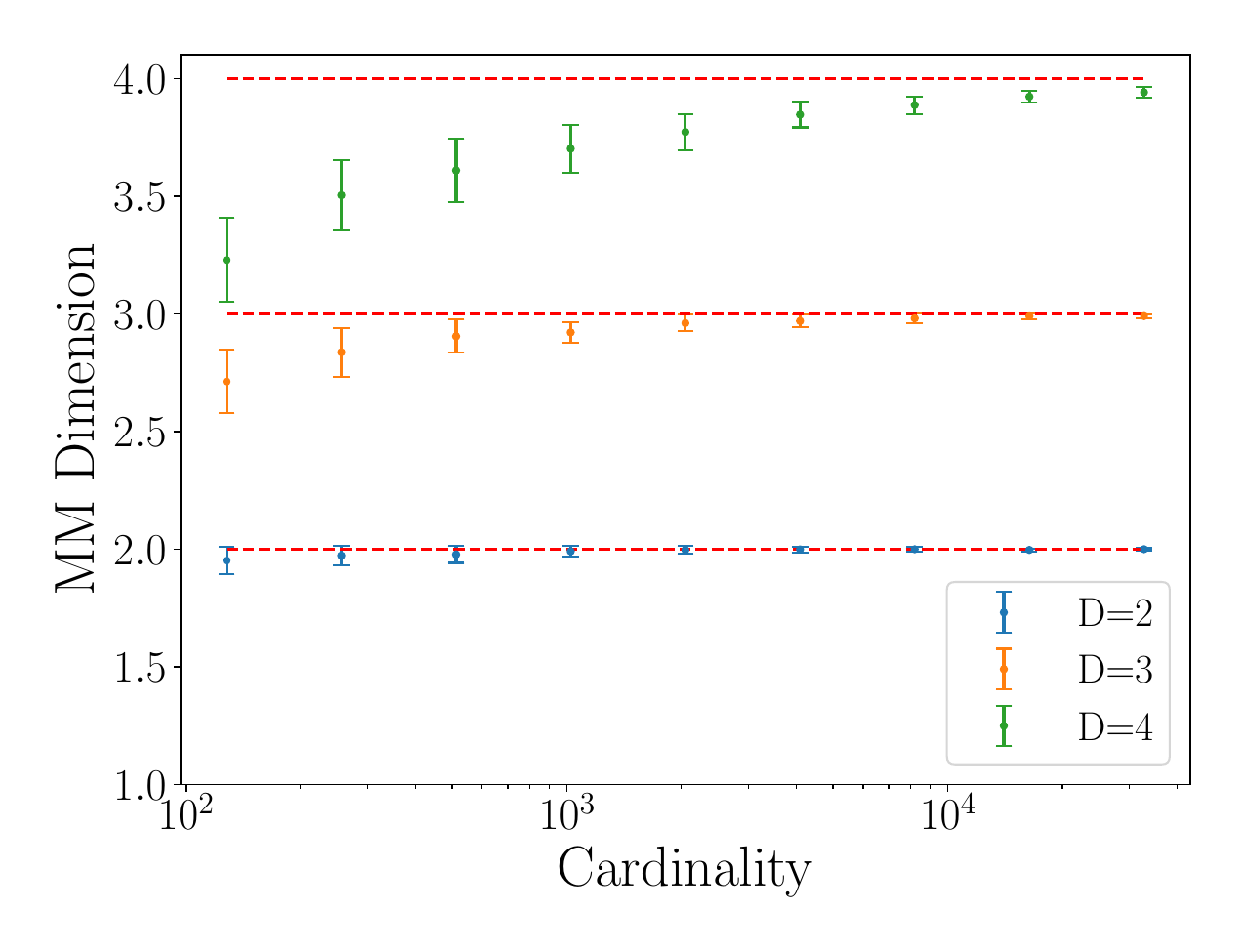}
    \vspace{-7mm}
    \caption{{Myrheim-Mayer dimension estimates converge to the dimensions of the Minkowski spacetime for $D=2,3,4$. Each data point is averaged over 200 causets.}}
    \label{fig: MMdim test}
\end{figure}

To further check that the causal structure in Minkowski spacetime works correctly we implemented a Myrheim-Mayer dimension estimator~\cite{Myrheim1978, Meyer1988}, which converges to the dimension of the spacetime that the causet is embedded in, as the size $N$ increases. 
Simulating causal sets up to $N=32768$, sprinkling into bicones in the dimensions $D=2,3,4$, we see that the dimension estimate agrees with the true value within the uncertainty, giving substantial evidence that our simulation framework works as intended (see Fig. \ref{fig: MMdim test}).

\begin{figure*}
    \phantomsubfloat{\label{fig:looser-bounds-results}}
    \phantomsubfloat{\label{fig:lambdas-3d-distribution}}
    \includegraphics[width=0.56\textwidth]{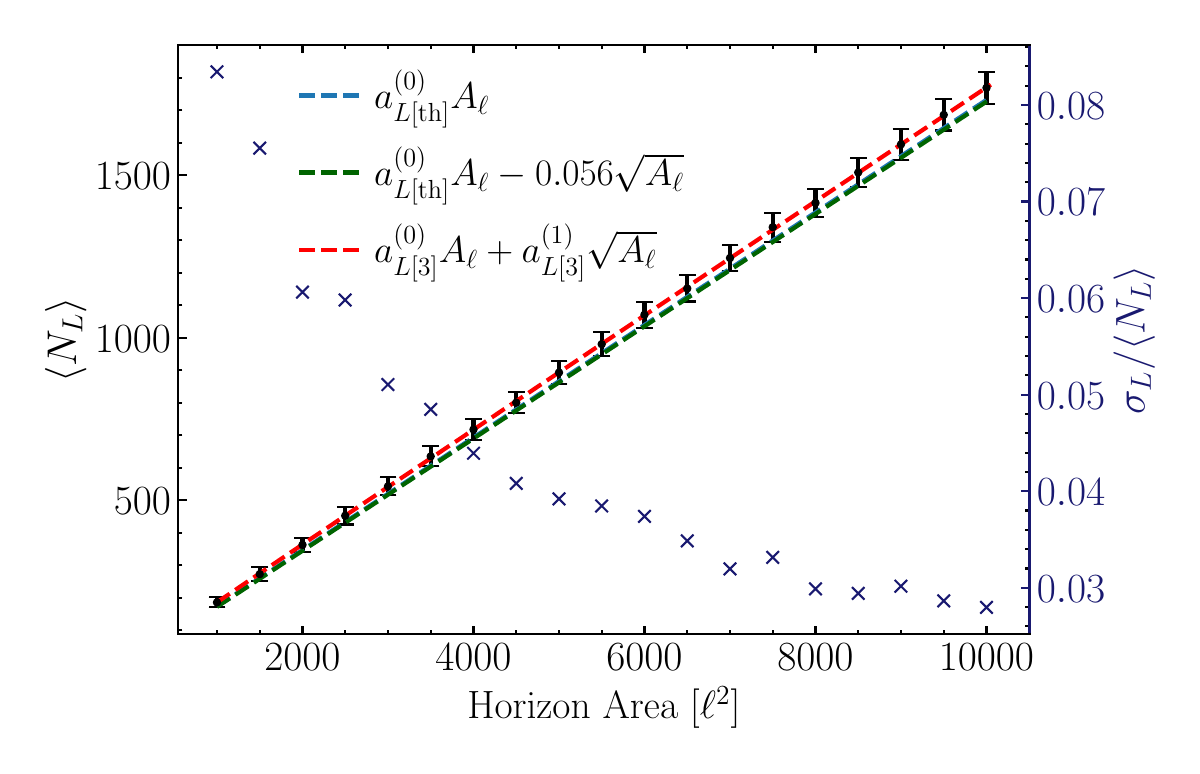}
    \begin{overpic}[width=0.41\textwidth]{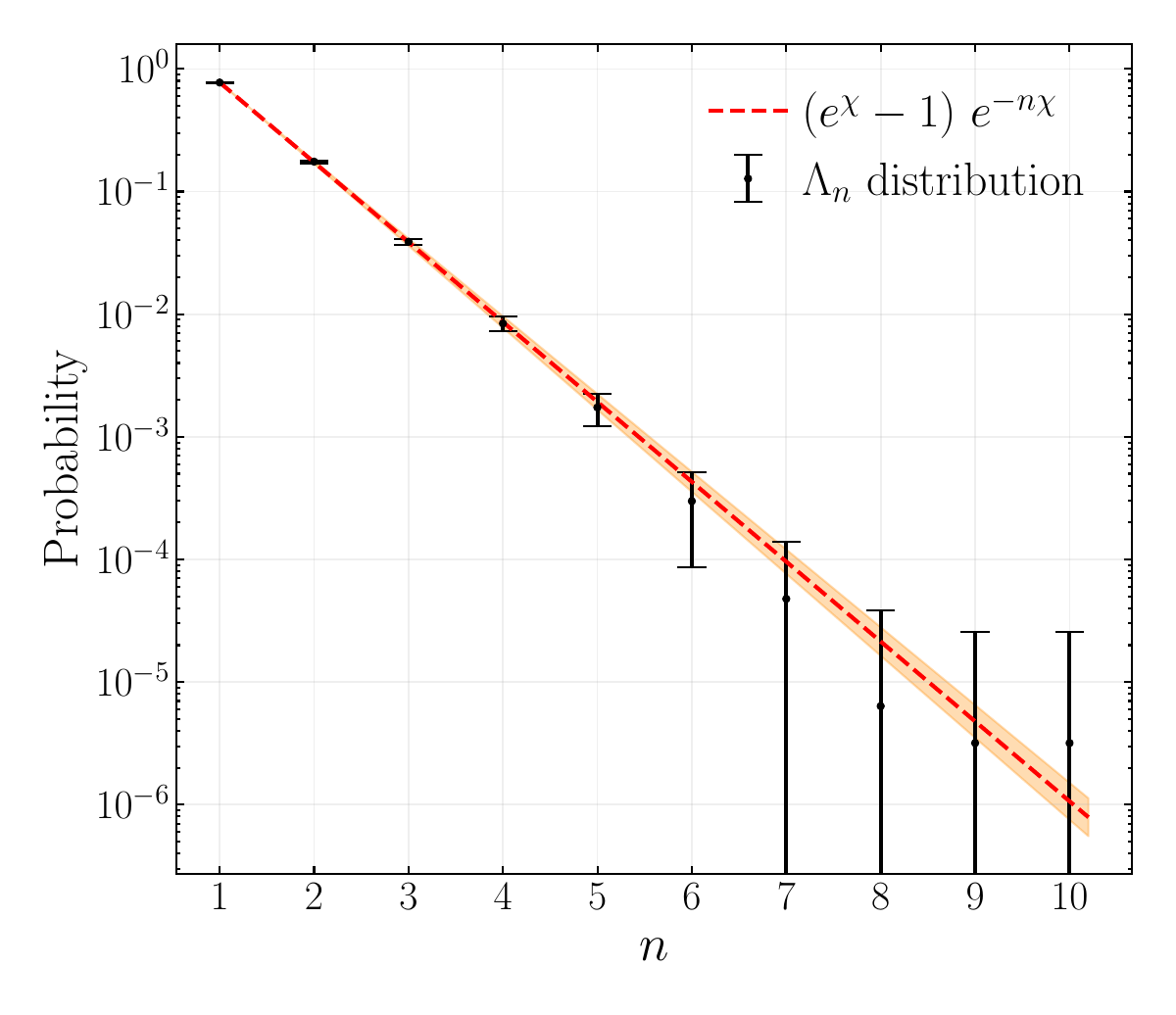}
        \put(-136, 88){(a)}
        \put(-2, 88){(b)}
    \end{overpic}
    \vspace{-7mm}
    \caption{(a) Counting of Link molecules from the simulations of causets in (3+1)D Schwarzschild spacetime, with simulations boundaries at distance $6\ell$ from relevant hypersurfaces $\Sigma$ and $\mathcal{H}$.
    Each datapoint corresponds to 100 simulations.
    The red line, with values from~\cref{eq:fit-with-wider-bounds}, is our best fit.
    The azure and green lines are the analytical models: respectively, the flat-spacetime approximation and the 1st-order curvature expansion.
    (b) $\Lambda_n$ molecules distribution in (2+1)D spacetime, from the simulation of 50 causets for each of 9 different area values in the range $A\in[500, 4500]\ell^2$.
    }   
\end{figure*}

\section{Other numerical results}\label{app:other-numerical-results}

In this appendix, we provide additional numerical results to the ones in \cref{Results}.

In the first place, we simulated causets in (3+1)D Schwarzschild spacetime in a region with bounds $t^* \in [-6, 0], \,r\in[r_S - 6,\, r_S + 6]$ in terms of discreteness units $\ell$, 
looser than the ones used for \cref{Results}.

The obtained results (see \cref{fig:looser-bounds-results}) are equivalent to those in \cref{Results}. 
Firstly, the analytical model lies within uncertainty, but is consistently lower than datapoints.
Secondly, the best fit is
\begin{equation}
    a^{(0)}_{L[\text{3}]}=0.173 \pm 0.005, \qquad a^{(1)}_{L[\text{3}]}=0.4 \pm 0.4,
    \label{eq:fit-with-wider-bounds}
\end{equation}
i.e. the same as in \cref{eq:best-fit}, but with larger uncertainties.
This provides an additional confirmation of the validity of the simulations in the main text.

Furthermore, we performed simulations in (2+1) dimensions, to investigate how the probability distribution $p_n$ of the horizon molecules varies in other dimensions.
The sprinkled region was bounded as the (3+1)D one in the main text: $t^* \in [-4, 0], \,r\in[r_S - 3,\, r_S + 3]$ in terms of discreteness units $\ell$.
We simulated 50 causets for each of 9 different equally-spaced horizon areas in the range $A\in[500, 4500]\ell^2$, with average cardinalities ranging roughly from $50,000$ to $450,000$.
We found that the molecules obey the same distribution as in (3+1)D, 
$p_n = \left(e^\chi - 1\right) e^{-n\chi}$,
with approximately the same coefficient
$\chi = 1.50 \pm 0.03$
(see \cref{fig:lambdas-3d-distribution}).
Therefore, the hypothesis outlined in \cref{subsec:results-discussion} that horizon molecules could satisfy some thermodynamic distribution is to be discarded.

\bibliography{refs}

\end{document}